\documentclass[sigconf]{acmart}
\usepackage[utf8]{inputenc}
\usepackage[export]{adjustbox}
\usepackage{subfigure}

\usepackage{threeparttable}
\usepackage{float}
\usepackage{makecell}
\usepackage{tikz}
\usetikzlibrary{arrows,positioning,decorations.pathreplacing}
\usepackage{url}
\usetikzlibrary{shapes}
\usepackage{xcolor}
\usepackage{balance}
\usepackage{xspace}
\usepackage{amsmath}

\usepackage{colortbl}
\usepackage{pgfplots} 

\definecolor{MinColor}{rgb}{1, 0.9, 0.9}  
\definecolor{MaxColor}{rgb}{0.6, 0, 0}  

\definecolor{lightorange}{HTML}{FFA07A}
\definecolor{lightblue}{HTML}{ADD8E6}

\hypersetup{
    colorlinks=true,
    linkcolor=blue,
    filecolor=magenta,      
    urlcolor=teal,
    }

\usepackage{pifont}
\usepackage{bbding} 
\newcommand{\cmark}{\textcolor{green!80!black}{\ding{51}}}
\newcommand{\xmark}{\textcolor{purple}{\ding{55}}}

\usepackage{cleveref}
\usepackage{amsfonts}%
\usepackage{multirow}
\usepackage{booktabs}
\usepackage{caption}
\usepackage{tcolorbox}
\usepackage{adjustbox}
\usepackage{array}

\usepackage{url}

\usepackage[ruled,vlined]{algorithm2e}
\SetAlFnt{\small}

\newcolumntype{R}[2]{%
    >{\adjustbox{angle=#1,lap=\width-(#2)}\bgroup}%
    c%
    <{\egroup}%
}

\usepackage[commandnameprefix=always]{changes}

\usepackage{todonotes}
\setuptodonotes{fancyline, inline, color=blue!30}
\usepackage[n,landau,notions,ff,mm]{cryptocode}

\definecolor{crg}{rgb}{0.5, 0.5, 0.1}   

\newcommand{\maxval}{30}     
\newcommand{\maxbarwidth}{1.1}

\newcommand{\cellbar}[1]{%
  \pgfmathparse{#1/\maxval*\maxbarwidth}%
  \edef\barlen{\pgfmathresult cm}%
  \begin{minipage}[c][2.5em][c]{\maxbarwidth cm}
    \colorbox{magenta!25}{\makebox[\barlen][l]{}}%
    \vspace{0.2em} \newline
    #1\%%
  \end{minipage}%
}

\usepackage{diagbox}

\usepackage[n,landau,notions,ff,mm]{cryptocode}
\definecolor{gamechangecolor}{gray}{0.74}

\usepackage{tikz}

\usepackage[framemethod=TikZ]{mdframed}
\mdfsetup{
    font=\small,
    leftmargin=5pt,
    rightmargin=5pt,
    skipbelow=5pt,
    skipabove=5pt,
    innertopmargin=5pt,
    innerbottommargin=5pt,
    innerleftmargin=5pt,
    innerrightmargin=5pt,
    frametitlerule=true,
    frametitlealignment=\hspace*{0pt}
}

\usepackage{subcaption}
\usepackage{float}

\usepackage{enumitem}

\usepackage{listings}

\definecolor{lightbrown}{RGB}{245, 235, 220}

\newenvironment{packeditemize}{
	\begin{list}{$\bullet$}{
			\setlength{\labelwidth}{4pt}
			\setlength{\itemsep}{0pt}
			\setlength{\leftmargin}{\labelwidth}
			\addtolength{\leftmargin}{\labelsep}
			\setlength{\parindent}{0pt}
			\setlength{\listparindent}{\parindent}
			\setlength{\parsep}{0pt}
			\setlength{\topsep}{1pt}}}{\end{list}}

\lstdefinestyle{custompython}{
    language=Python,
    basicstyle=\ttfamily\small,
    escapeinside={(*@}{@*)},  
    commentstyle=\color{gray}\itshape,
    keywordstyle=\color{blue!70!black}\bfseries,
    stringstyle=\color{orange!70!black},
    showstringspaces=false,
    breaklines=true,
    frame=single,
    columns=flexible
}

\newcommand{\hlhref}[2]{\href{#1}{\textcolor{teal}{{#2}}}}

\usepackage{fontawesome}
\usepackage{acro}
\acsetup{single}

\DeclareAcronym{IDM}{
  short = IDM,
  long  = Input Data Message,
}
\newcommand{\IDM}{\ac{IDM}\xspace}
\newcommand{\IDMs}{\acp{IDM}\xspace}

\DeclareAcronym{EOA}{
  short = EOA,
  long  = Externally Owned account,
}
\newcommand{\EOA}{\ac{EOA}\xspace}
\newcommand{\EOAs}{\acp{EOA}\xspace}

\newcommand{\inputdata}{\texttt{input} \texttt{data}\xspace}

\newcommand{\StartTimeData}{Jul~$30$,~$2015$\xspace}
\newcommand{\EndTimeData}{Feb~$26$,~$2024$\xspace}

\newcommand{\EndBlockData}{19{,}314{,}987\xspace}

\newcommand{\numDecodableTXs}{5{,}238{,}336\xspace}

\newcommand{\numMeaningfulTXs}{867{,}140\xspace}
\newcommand{\numMeaningfulTXStructOnly}{444{,}753\xspace}
\newcommand{\numMeaningfulTXsLangOnly}{366{,}338\xspace}
\newcommand{\numMeaningfulTXsMixed}{56{,}049\xspace}

\newcommand{\numMeaningfulTXsWithLangNotOnly}{422{,}387\xspace}

\newcommand{\numEnglishIDMTotal}{402{,}801\xspace}
\newcommand{\pctEnglishIDM}{$95.4$\%\xspace}

\newcommand{\numChineseIDMTotal}{18{,}557\xspace}
\newcommand{\pctChineseIDM}{$4.4$\%\xspace}

\newcommand{\numOtherLangIDMTotal}{1{,}029\xspace}
\newcommand{\pctOtherLangIDM}{$0.2$\%\xspace}

\newcommand{\numIDMFromAddr}{59{,}795\xspace}

\newcommand{\numIDMToAddr}{154{,}411\xspace}

\newcommand{\numENIDMwithEmotion}{269{,}445\xspace}
\newcommand{\numCNIDMwithEmotion}{15{,}370\xspace}

\newcommand{\TotalWithNaturalTextTXs}{\ensuremath{422{,}387}\xspace}

\newcommand{\UniqueWithNaturalTextTXs}{\ensuremath{60{,}847}\xspace}

\newcommand{\TotalAddrsWithNaturalTextTXs}
{\ensuremath{189{,}111}\xspace}

\newcommand{\ConnectedAddrs}{\ensuremath{189{,}006}\xspace}

\newcommand{\SelfSentAddrs}{\ensuremath{5{,}230}\xspace}

\newcommand{\SelfSentTxs}{\ensuremath{14{,}185}\xspace}

\newcommand{\numOfCommunity}{\ensuremath{26{,}048}\xspace}

\newcommand{\ETH}{\texttt{ETH}\xspace}

\renewcommand\footnotetextcopyrightpermission[1]{} %
\begin{document}

\makeatletter
\def\@mkfootnotetext#1{\parindent 1em\noindent\hbox to 1.8em{\hss$^{\textcolor{green}{*}}$}#1\par}
\makeatother

\title{Talking Transactions: Decentralized Communication through Ethereum Input Data Messages (IDMs)}

\author{Xihan Xiong}
\affiliation{
  \institution{\textit{Imperial College London}}
}

\author{Zhipeng Wang\textsuperscript{\scriptsize \faEnvelopeO}
}
\affiliation{
      \institution{\textit{Imperial College London}}
}

\author{Qin Wang}
\affiliation{
  \institution{\textit{CSIRO Data61}}
}

\author{Endong Liu}
\affiliation{
  \institution{\textit{University of Birmingham}}
}

\author{Pascal Berrang}
\affiliation{
  \institution{\textit{University of Birmingham}}
}

\author{William Knottenbelt}
\affiliation{
  \institution{\textit{Imperial College London}}
}

\begin{abstract}

Can you imagine, \textit{\textbf{blockchain transactions can talk!}} In this paper, we empirically study how they talk and what they talk about. 

We focus on the \inputdata field of Ethereum transactions, which is designed to allow external callers to interact with smart contracts. In practice, this field also enables users to embed natural language messages into transactions.  Users can leverage these Input Data Messages (IDMs) for peer-to-peer communication. This means that, beyond Ethereum's well-known role as a financial infrastructure, it also serves as a decentralized communication medium.

We present the first large-scale and systematic analysis of Ethereum IDMs from the genesis block to February 2024 (3134 days, covering 87\%+ historical transactions to date).  
We filter IDMs to extract \numMeaningfulTXs transactions with informative text messages and use LLMs for language detection. We find that English (\pctEnglishIDM) and Chinese (\pctChineseIDM) dominate the use of natural languages in IDMs.
Interestingly, English IDMs center on security and scam warnings ($24\%$) with predominantly negative emotions, while Chinese IDMs emphasize emotional expression and social connection ($44\%$) with a more positive tone.
We also observe that longer English IDMs often transfer high ETH values for protocol-level purposes, while longer Chinese IDMs tend to involve symbolic transfer amounts for emotional intent.
Moreover, we find that, unlike traditional social networks, the IDM participants tend to form small, loosely connected communities ($59.99\%$). 
Our findings highlight culturally and functionally divergent use cases of the IDM channel across user communities.

We further examine the security relevance of IDMs in on-chain attacks. Many victims use them to appeal to attackers for fund recovery. IDMs containing negotiations or reward offers are linked to higher reply rates.
We also analyze IDMs' moderation and regulation implications. Their misuse for abuse, threats, and sexual solicitation reveals the urgent need for content moderation, regulation, and governance in decentralized systems.

\end{abstract}

\maketitle

{
\makeatletter
\renewcommand{\@makefnmark}{\hbox{\@textsuperscript{\faEnvelopeO}}}
\makeatother
\footnotetext{Corresponding author, \href{mailto:zhipeng.wang0x01@gmail.com}{zhipeng.wang0x01@gmail.com}.}
}


\section{Introduction}
The Bitcoin whitepaper~\cite{bitcoin} originally framed blockchain technology as a peer-to-peer \emph{electronic payment system}. This foundational vision was significantly extended by Ethereum~\cite{wood2014ethereum}, which is a quasi-Turing-complete blockchain that enables smart contracts. Smart contracts are self-executing programs that operate on the blockchain when predefined conditions are met. Ethereum allows users to interact with these contracts by specifying parameters in the \inputdata field of transactions, thereby supporting a wide range of decentralized applications beyond simple payments~\cite{antonopoulos2018mastering}.

Interestingly, in practice, this \inputdata field can be repurposed by users to embed arbitrary natural language messages directly into the transactions. This allows the Ethereum blockchain to function as a decentralized communication medium. Much like traditional messaging platforms, users can send messages, convey opinions, and exchange 
 information via \IDMs~\cite{idm2023}.

\begin{center}
   \colorbox{cyan!7}{
    \begin{minipage}{0.9\linewidth}
        A notable example is the Bybit Exploiter~\cite{bybit2025}, one of the most notorious actors in blockchain history, who stole over $1.4$b USD from Bybit in February 2025. Since then, the exploiter address received more than \href{https://etherscan.io/idm?addresses=0x47666fab8bd0ac7003bce3f5c3585383f09486e2\%2C}{$4{,}100$ IDMs} from the public, where angry victims demand the return of stolen funds and opportunistic bystanders plead for a share of the stolen assets.
    \end{minipage}
} \end{center}

The Etherem \IDM volume has increased significantly since $2018$ (see Figure~\ref{fig:number_of_txs_and_idms}). 
Blockchain users' active adoption of IDMs signals a paradigm shift -- blockchain is no longer merely a ``value internet'' where users engage in financial activities, but also a socio-technical infrastructure for users to talk in a decentralized manner. Although \IDMs have been introduced and widely used since the inception of Ethereum in $2015$, this beyond-finance functionality of blockchains has remained underexplored in academia during the last decade.

To address this gap, this paper presents the first large-scale and systematic study of Ethereum \IDMs (see Figure~\ref{fig: Pipeline}). By adopting a ``transaction-as-communication'' perspective, we empirically analyze how users are talking and what they are talking about. 
We summarize our main contributions as follows:

\begin{figure*}[t]
\centering
\includegraphics[width=0.9\linewidth]{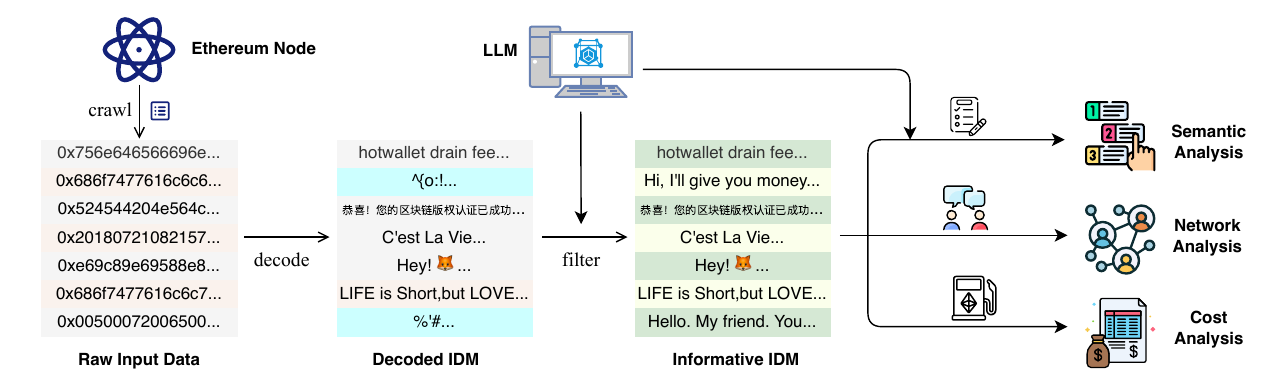}
\vspace{-0.1in}
\caption{Pipeline of Ethereum \IDM analysis.}
\label{fig: Pipeline}
\vspace{-0.1in}
\end{figure*}

\begin{packeditemize}
    \item \textbf{Large-Scale IDMs Analysis } (\S\ref{sec:data} and \S\ref{sec: descriptive}). 
    We conduct the first large-scale analysis of Ethereum \IDMs from the inception of Ethereum in July 2015 to February 2024, covering 87\%+ of historical transactions as of May 2025.  We then filter IDMs to extract \numMeaningfulTXs transactions with informative text messages. These involve \numIDMFromAddr senders and \numIDMToAddr receivers. We find that these IDMs occupy $0.12$~GB of an Ethereum full node storage. Our descriptive analysis shows that $51.3\%$ of IDMs contain only structured tokens (e.g., wallet address), $42.2\%$ contain only natural language, and $6.5\%$ contain both. We find that Chinese and English together represent $99.8\%$ of all \IDMs with natural languages.

    \vspace{0.5mm}
    \item \textbf{Semantic Analysis} (\S\ref{sec: semantic}).  
    Our topic analysis identifies 12 main topics with 48 subtopics. We find that English \IDMs are concentrated in \textit{Security \& Incidents} ($24\%$), while Chinese \IDMs center on \textit{Social \& Emotional Expression} ($44\%$). For sentiment analysis, we develop a taxonomy with three polarities and 16 emotion categories. We discover that negative emotions (e.g., \textit{Fear}) dominate the English IDMs, while positive emotions  (e.g., \textit{Joy}) dominate Chinese \IDMs. This reflects two contrasting communicative logics: one driven by risk and alert, the other by social presence.

   \vspace{0.5mm}
    \item \textbf{Cost Analysis} (\S\ref{sec-idmComm}). 
    We analyze \IDM transaction value and cost.  We find that longer English IDMs are often associated with higher ETH transfers for functional or protocol-level purposes. In contrast, longer Chinese IDMs tend to involve symbolic transfer amounts (e.g., 5.20 ETH) for emotional intent. We also find that gas costs per byte fell sharply after EIP-2028, yet \IDM volume did not rise until mid-2023. This suggests that IDM adoption is more likely driven by social usage than by cost efficiency.

    \item \textbf{Network Analysis} (\S\ref{sec-idmNetwork}).  
    We conduct a network analysis of IDM participants and identify \numOfCommunity communities.  Most communities ($59.99\%$) are small and loosely connected, with low reciprocity and clustering coefficient. Only a few large communities dominate message traffic, often driven by promotional or warning-oriented broadcasters. The largest community accounts for $34.9\%$ of all IDMs issued. In contrast, small communities prefer self-expression and social connection. This highlights the coexistence of information hubs and peripheral emotional expression.

    \vspace{0.5mm}
    \item \textbf{Security Relevance} (\S\ref{sec-security}). 
    We examine the relevance of \IDMs to on-chain security incidents. Many victims use IDMs to reach out to attackers: pleading for fund returns, issuing threats, offering rewards, or proposing negotiations. Our analysis suggests that negotiation and reward offers are more likely to increase the reply rate. Beyond victim communication, \IDMs are also used to broadcast warnings about scams. This suggests that IDMs help enable collective safeguarding in the Ethereum community.

    \vspace{0.5mm}
    \item \textbf{Regulation and Moderation Implications} (\S\ref{sec-regulation}).
    We analyze IDMs classified under the topic of \textit{Toxic/Abusive Content}. English IDMs contain significantly more toxic content than Chinese ones, particularly concentrated in \textit{Verbal Abuse \& Profanity}. Most of these messages are linked to negative emotions, especially \textit{Hostility} and \textit{Anger}. Our findings reveal that blockchain messaging can be misused for severe forms of abuse and harassment. While such content would typically be removed in Web2 platforms through content moderation systems, blockchain ecosystems lack such mechanisms. This underscores the need for regulatory attention and content governance in decentralized environments.
    
\end{packeditemize}

\section{System Model}
\label{sec-model}

\noindent\textbf{\IDM Transaction Metadata.}
We consider the transactions in Ethereum, a Virtual Machine (EVM)-compatible blockchain. A standard Ethereum transaction includes a variety of fields that enable both coin transfers and smart contract interactions. We list the transaction fields that are relevant to \IDMs analysis as follows:

\begin{packeditemize}
    \item \emph{Transaction Hash}: A 32-byte unique ID derived from a transaction’s contents, used for referencing or retrieval.
    \item \emph{Block Number}: The height of the block in which the transaction was confirmed and executed. This provides the timestamp for events or messages in the transaction.
    \item \emph{From Address}: The \EOA that initiates the transaction. This address is the sender of a transaction and is responsible for paying the gas fees to execute the transaction. 
    \item \emph{To Address}: The recipient address, which may be either another \EOA or a smart contract. In the context of IDMs, we focus solely on cases where the recipient address is an \EOA.
    \item \emph{Value}: The amount of \ETH transferred from the transaction sender to the recipient when the transaction is confirmed. 
    
    \item \emph{Input Data}: The \inputdata is an optional field that enables encoding function calls and parameters for smart contracts. In our context, it is often repurposed by \IDM senders to embed human-readable messages in hexadecimal or UTF-8 format.
\end{packeditemize}

\vspace{0.6mm}
\noindent\textbf{\IDM Transaction Workflow}. To issue a transaction containing an \IDM, the sender follows the same process as sending a typical Ethereum transaction, with a key distinction: instead of adding the parameters in the \inputdata field to interact with a smart contract, the sender includes a message. This \IDM is typically formatted in hexadecimal or UTF-8 encoding and can be decoded to human-readable messages. The transaction is then broadcast to the network, executed and confirmed by validators, and permanently recorded on the blockchain once included in a block. The transaction recipient, and anyone who has access to the blockchain data, can extract and decode the \IDM content from the transaction \inputdata.

\vspace{0.6mm}
\noindent\textbf{Challenges of \IDM Analysis.}
Ethereum \IDMs present unique analytical challenges due to the following features.
\begin{packeditemize}
    \item \emph{Domain-specific}: 
    IDM content often reflects blockchain-specific concepts or operations. However, general-purpose NLP tools are not trained to handle such specialized vocabulary and syntax.
    \item \emph{Mixed modality}:
    IDMs often embed structured tokens (e.g., wallet addresses, see \S\ref{sec:data}) within natural languages. These heterogeneous inputs complicate tokenization and downstream semantic tasks.
    \item \emph{Code-switching}: 
    Some IDMs mix languages within a single message (e.g., English with Chinese). This challenges monolingual models and language-specific pre-processing pipelines.
    \item \emph{Noisy}:
    IDMs are often ungrammatical, abbreviated, or informal. They may include typos, slang, and inconsistent punctuation. This noise reduces the effectiveness of traditional NLP tools.

\end{packeditemize}

\vspace{0.6mm}
\noindent\textbf{LLM-assisted Analysis.} 
Given IDMs' unique features and analytical challenges, we use an LLM, \texttt{GPT-4o}, to assist several stages of our analysis. First, it helps refine language detection (\S\ref{sec:data}). Second, we use the LLM to assign IDM topics based on a predefined taxonomy (\S\ref{sec: topic_analysis}). It also identifies the emotional tone of each message using a structured set of sentiment categories (\S\ref{sec: sentiment_analysis}). Finally, we apply the LLM to classify fund recovery requests by strategy type (\S\ref{sec: idm_security}). An example LLM prompt is provided in Appendix~\ref{appendix: prompt}.

\section{Data Collection and Preprocessing}
\label{sec:data}

We first crawl Ethereum transactions within the time frame of our empirical analysis. We then extract a decodable IDM dataset that includes transactions whose \inputdata can be decoded using UTF-8. Finally, we filter this set to obtain an informative IDM dataset.

\smallskip
\noindent\textbf{Raw Dataset.} We collect Ethereum transactions from the genesis block (\hlhref{https://etherscan.io/block/1}{Block\#0}) to \hlhref
{https://etherscan.io/block/19314987}{Block\#\EndBlockData} ($3{,}132$ days, from \StartTimeData to \EndTimeData). This covers the majority of Ethereum’s history up to a recent snapshot (over \textbf{87\%} as of May 2025). 

For each transaction, we extracted the \inputdata field. In transactions where an \EOA interacts with a smart contract, this field typically contains ABI-encoded function names and parameters for contract execution. The semantic content of such data is weak in terms of human interpretability, as its structure and meaning are defined by contract logic rather than natural language. Therefore, our focus is on transactions between two \EOAs, where the input data is unconstrained by ABI standards and may carry natural language content beyond basic transaction metadata.

\smallskip
\noindent\textbf{Decodable IDM Dataset.} 
To identify text messages embedded in the \inputdata field, we attempted to decode the \inputdata using \texttt{UTF-8} encoding. If the decoding was successful, we treated the transaction as a candidate for \IDMs. For each candidate transaction, we stored its transaction hash, block number, block timestamp, from and to addresses, gas used, gas limit, gas price, original hexadecimal input data, input data length, and decoded UTF-8 text. In total, we have identified \numDecodableTXs transactions with decodable \IDMs, with a total data size of $480{,}932{,}992$ bytes ($\approx 0.48$~GB).

\smallskip
\noindent\textbf{Informative IDM Dataset.}
After collecting all \texttt{UTF-8} decodable IDM transactions, we performed a filtering step to identify messages that carry meaningful information. We first extracted structured tokens from the decoded text. These included URLs, wallet addresses, transaction hashes, emojis, references to on-chain operations (e.g., cross-chain bridging, exchange operations), and embedded Base64-encoded media (e.g., images in JPEG formats). 

Then we use \texttt{FastText} for language detection. We cleaned the decoded text by removing structured tokens. This ensured that the \texttt{FastText} model operated on the remaining natural language content.  
However, \texttt{FastText} exhibits limitations when applied to domain-specific messages that contain noisy, mixed-modality, and code-switching text~\cite{jauhiainen2019automatic}. Therefore, we used \texttt{GPT-4o} as a secondary step to refine IDM language detection (see Figure~\ref{fig: Pipeline}). The detected language was then stored for further analysis.  

A transaction was considered to contain an informative \IDM if it met either of the two criteria: \emph{(i)} the text contained any recognizable structured tokens, regardless of whether natural language was present; \emph{(ii)} if the text contained human-readable natural language, even without structured tokens. Transaction text that contained neither was considered as noise and discarded. 

After this, we obtain a dataset with informative IDMs. The size of the IDM data is $121{,}237{,}387$ bytes ($\approx$ $0.12$~GB).

\section{Descriptive Analysis}
\label{sec: descriptive}
This section provides a descriptive analysis of our \IDM dataset.

\begin{table}[tbh]
\centering
\caption{IDM types by content and language variety.}
\label{tab: IDM_type_by_lang_and_content}
\vspace{-0.5em}
\resizebox{\columnwidth}{!}{
\begin{tabular}{c|ccccc}
\toprule
\multicolumn{1}{c}{\textbf{IDM Type}} & \textbf{Contain NL} & \textbf{Contain ST} & \multicolumn{1}{c}{\textbf{Language Variety}} & \multicolumn{1}{c}{\textbf{Count}} & \textbf{Total} \\
\midrule
 ST Only & \xmark & \cmark & None & \cellcolor{gray!0} 444,753 & 444,753 \\
 
\cmidrule{4-6}

\multirow{2}{*}{NL Only} 
                           & \multirow{2}{*}{\cmark} 
                           & \multirow{2}{*}{\xmark} 
                           & Monolingual & \cellcolor{gray!0} 365,529  & \multirow{2}{*}{366,338} \\
                           & & &  Multilingual & \cellcolor{gray!0}  809 & \\
\cmidrule{4-6}

\multirow{2}{*}{Mixed Type} 
                           & \multirow{2}{*}{\cmark} & \multirow{2}{*}{\cmark} & Monolingual  & \cellcolor{gray!0} 50,963   & \multirow{2}{*}{56,049} \\ 
                           & & &  Multilingual & \cellcolor{gray!0} 5,086    &\\
\bottomrule

\end{tabular}
}

\begin{tablenotes}[flushleft]
      \footnotesize
      \item[] \quad \textit{\textbf{NL} for natural language; \textbf{ST} for structured token.}
\end{tablenotes}
\end{table}

After data preprocessing, we obtain \numMeaningfulTXs transactions with informative \IDMs. Among these, \numMeaningfulTXStructOnly transactions (51.3\%) contain only structured tokens, \numMeaningfulTXsLangOnly transactions (42.2\%) contain only natural languages, and \numMeaningfulTXsMixed transactions (6.5\%) contain both (see Table~\ref{tab: IDM_type_by_lang_and_content}). 
The composition of informative IDMs reveals that structured token dominates. This suggests that many on-chain messages lack explicit communicative intent. Nevertheless, a substantial portion contains natural language (\numMeaningfulTXsWithLangNotOnly, 48.7\%). This indicates that users do leverage transactions for linguistic expression.

\begin{table}[tbh]
    \centering
    \caption{Top languages in IDMs and their usage types.}
    \vspace{-0.08in}
    \label{tab:language_distribution}
    \resizebox{\columnwidth}{!}{
    \begin{tabular}{c|cc|c|cc}
    \toprule
    \multicolumn{1}{c}{\textbf{Language}} & \# Monolingual & \multicolumn{1}{c}{\# Multilingual} & \multicolumn{1}{c}{\textbf{Language}} & \# Monolingual & \# Multilingual \\
    \midrule
    English & 396{,}953 & 5848 & Chinese & 17{,}731 & 826 \\
    German & 222 & 44 & Korean & 212 & 41\\
    French & 186& 17 & Japanese & 157 & 73\\
    Spanish &142 &18  &Russian& 106& 75\\
    \bottomrule
    \end{tabular}
    }
    \begin{tablenotes}[flushleft]
      \footnotesize
      \item[] \quad \textit{\textbf{Monolingual} indicates IDMs containing only one language; \textbf{Multilingual} indicates IDMs containing the target language as part of a multilingual message.}
    \end{tablenotes}
\end{table}

\smallskip
\noindent\textbf{Top Languages}.
Table~\ref{tab:language_distribution} shows the most common languages in \IDMs and their distribution across monolingual and multilingual contexts. English dominates the linguistic content of \IDMs, appearing in \numEnglishIDMTotal instances (\pctEnglishIDM). Chinese appears as the second most frequent language (\numChineseIDMTotal, \pctChineseIDM), though with a substantially lower count. Together, English and Chinese are present in 99.8\% of the IDMs that contain natural languages.  
Other languages (\numOtherLangIDMTotal, \pctOtherLangIDM), such as German, French, Spanish, Korean, Japanese, and Russian, occur at much smaller scales, typically in the hundreds. Across all listed languages, the number of \IDMs containing the language as a monolingual message consistently exceeds the number of instances where the language is part of a multilingual message.

\begin{figure}[htb]
\centering
\includegraphics[width=\columnwidth]{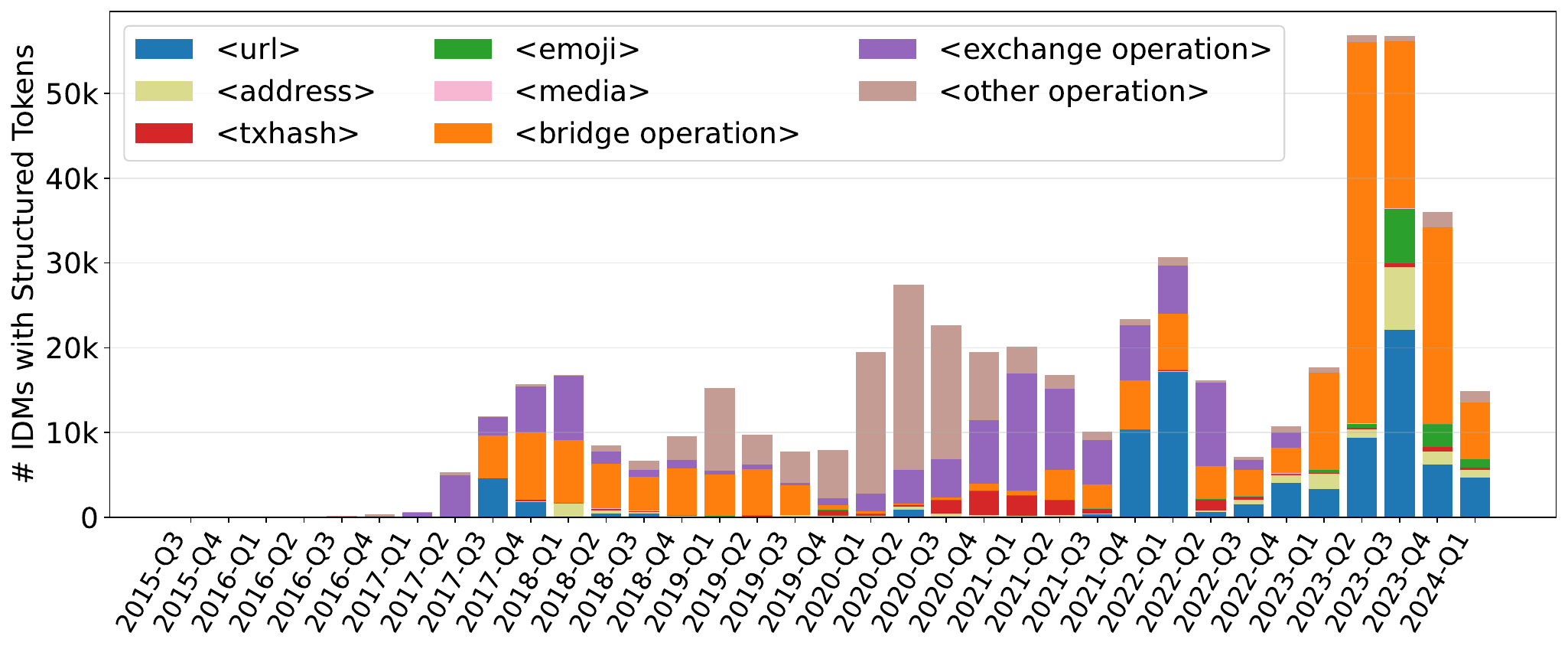}
\caption{Temporal trends of IDM with structured tokens.}
\label{fig:structured_tokens}
\end{figure}

\smallskip
\noindent\textbf{Structured Tokens}.
Figure~\ref{fig:structured_tokens} illustrates the evolution of structured tokens embedded in Ethereum \IDMs. 
The sharp increase of the URL element in $2023$ likely reflects the rise of Web3 project promotions and community invitations (e.g., Telegram links) embedded directly in on-chain messages. These promotional and invitational messages often also include emojis to enhance visibility and engagement.
The spike in exchange operation elements in $2021$ may be attributed to the surge in DEX activity and the increased use of on-chain aggregators and trading bots, many of which embed structured indicators or routing metadata directly within transaction messages.
In addition, bridge-related operations saw a significant increase in $2023$. This trend is likely driven by the growing adoption of on-chain interoperability and cross-chain DeFi protocols~\cite{werner2022sok,jiang2023decentralized}.

\begin{figure}[htb]
\centering
\includegraphics[width=\columnwidth]{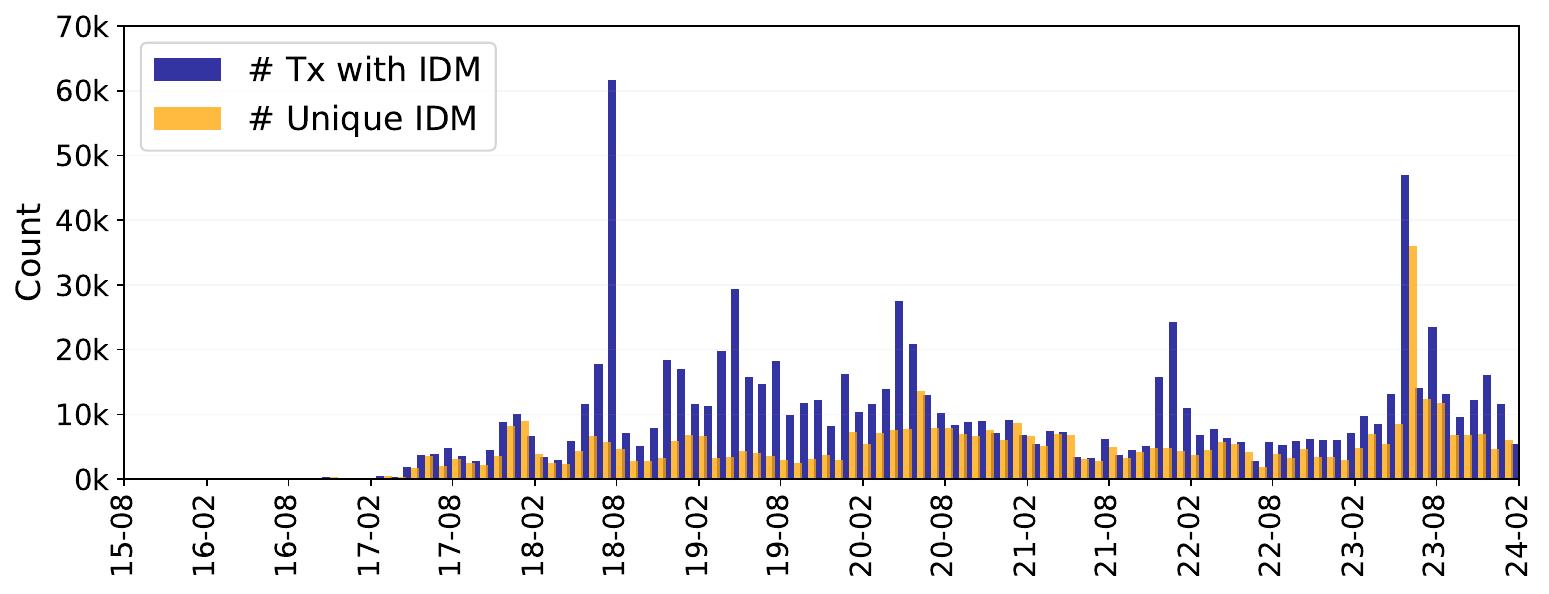}
\caption{Temporal trends of transactions with IDMs.}
\label{fig:number_of_txs_and_idms}
\end{figure}

\smallskip
\noindent\textbf{IDM Volume Trends}. Figure~\ref{fig:number_of_txs_and_idms} illustrates the monthly trends of Ethereum transactions containing \IDMs and the number of unique \IDMs from 2015 to 2024. 
Early in the timeline, the on-chain messaging activity was very low. Before May~$2017$, fewer than $1{,}000$ transactions per month contained \IDMs. Starting in the second half of $2017$, the volume began to rise steadily.  A major spike is observed in August 2018, with $61{,}678$ transactions containing \IDMs. However, only $4{,}595$ of these are unique, suggesting a high degree of repetition. 
After $2018$, activity drops but remains lively, with smaller peaks at irregular intervals. 
In general, IDM usage on Ethereum follows no fixed cycle. Its volume may be influenced by various events, such as social movements or security incidents.

\begin{figure}[htb]
\centering
\includegraphics[width=\columnwidth]{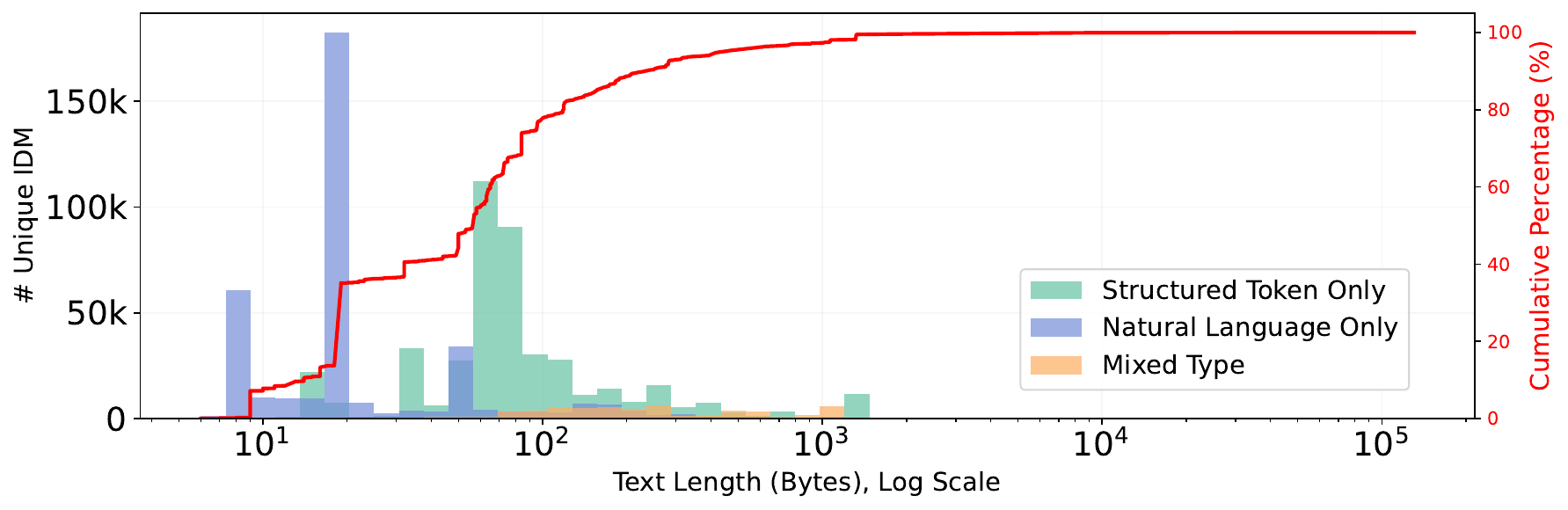}
\caption{Text length distribution by IDM types.}
\label{fig:lnput_len_distribution}
\end{figure}

\smallskip
\noindent\textbf{IDM Text Length}. 
Figure~\ref{fig:lnput_len_distribution} shows the distribution of unique \IDMs by text length. The distribution is highly skewed with a long-tail pattern. Most \IDMs are short. The majority fall between 10 and 100 bytes. This suggests that many messages are concise.  About $80\%$ of all unique IDMs are shorter than 100 bytes. Very few exceed $1{,}000$ bytes. These long messages likely contain encoded data or media.

Furthermore, \IDMs that contain only structured tokens (green) tend to concentrate around $60$ to $80$ bytes. In contrast, natural language messages (blue) are concentrated at shorter lengths, often $20$ bytes. Mixed-type messages (orange) are relatively sparse but more evenly distributed across the mid-range.


\begin{table*}[ht]
\centering
\caption{Overview of the Ethereum IDM topic taxonomy.} 
\label{tab:topic_taxonomy}
\vspace{-0.5em}
\renewcommand\arraystretch{1}
\resizebox{\linewidth}{!}{
\begin{tabular}{c|c|ll|c|cccc}
\toprule

\multicolumn{1}{c}{\multirow{2}{*}{\textbf{Main Topic}}} & \multicolumn{1}{c}{\multirow{2}{*}{\makecell{\textbf{Percentage}}}}  & \multicolumn{1}{c}{\multirow{2}{*}{\textbf{Subtopic}}} & \multicolumn{1}{c}{\multirow{2}{*}{\textbf{Description}}} & \multirow{2}{*}{\textbf{Example}} & \multicolumn{2}{c}{\textbf{English}} & \multicolumn{2}{c}{\textbf{Chinese}} \\

\cmidrule(lr){6-7} \cmidrule(lr){8-9}

\multicolumn{2}{c}{} & & \multicolumn{1}{c}{} & & \textbf{\#Unique}  & \textbf{\#Total} & \textbf{\#Unique} & \textbf{\#Total} \\
 
\midrule

\multicolumn{1}{c|}{\multirow{9}{*}{\makecell{\textbf{Social \&}\\ \textbf{Emotional} \\ \textbf{Expression}}} } & \multirow{9}{*}{\cellbar{22.3}} & Love/Confession & Love declarations or proposals. & \hlhref{https://etherscan.io/tx/0x10db14f2a83e139f30cc2114f4441e1baead0beaaab0e46ae7c7dc33a53f8a57}{0x10d...a57} & 581 & 652 & 2,459 & 5,313 \\

& & Despair & Expressions of emotional breakdown. & \hlhref{https://etherscan.io/tx/0x8da2d313d35b3a139fd32933f8560374fb9295d03027e4758dab7872c1018d83}{0x8da...d83} & 1,806 & 2,054 & 453 & 462 \\
 
& & Daily Life Record  & Sharing personal life moments or experiences. & \hlhref{https://etherscan.io/tx/0x6d7164ee3b7dc9faa2df027558967a5aa51be4f3d07fd368a464701a6eb2afc7}{0x6d7...fc7} & 1,012 & 1,114 & 1,070 & 1,111 \\
 
& & Greetings & Polite message for welcome or recognition. & \hlhref{https://etherscan.io/tx/0x0afeec928e6c21fe89b91fffb37de6ac724ab3bee7aed2bb708f833a5677462f}{0x0af...62f} & 1,735 & 3,925 & 234 & 278 \\
 
& & Birthday/Celebration & Birthday wishes, holidays, celebration. & \hlhref{https://etherscan.io/tx/0xd94d409337c25771ec5ad290cb42447b0c56487b076ce4c4e3aadc3189898349}{0xd94...349} & 522 & 589 & 984 & 1,156 \\
 
& & Philosophical Expression & Philosophical statements. & \hlhref{https://etherscan.io/tx/0x4c83978a6e7c0d0fdfb81245a40839cdddb210c59fb407e8f4c853c88d383d84}{0x4c8...d84} & 678 & 861 & 624 & 670 \\
 
& & Memorial/Tribute & Tributes to people or events. & \hlhref{https://etherscan.io/tx/0x6c5d6feb9a79bd40e38edf9b0e784f50a8f26d4f857418bad3ab647cc12ed02c}{0x6c5...02c} & 712 & 863 & 152 & 165 \\
 
& & Encouragement/Gratitude & Encouragement, motivation, or gratitude. & \hlhref{https://etherscan.io/tx/0x341c2f5db1d26ee7ea17cbfc4f22d10fa6756c768b1098723112cbe4a16d9645}{0x341...645} & 94 & 102 & 3 & 3 \\
 
& & Social Interaction & Messages intended to initiate informal interpersonal contact. & \hlhref{https://etherscan.io/tx/0x56eb823e43657aae5a00ec81a72906f27b21de00621bcdbdd404b942047ac943}{0x56e...943} & 36 & 49 & 13 & 14 \\
 
\midrule

\multirow{3}{*}{\makecell{ \textbf{Security \& } \\ \textbf{Incident} }} &  \multirow{3}{*}{\cellbar{18.7}} & Warnings & Warnings on malicious activities, e.g., phishing, scams, rug pulls. & \hlhref{https://etherscan.io/tx/0x14dc72531b0a442b246fc514fe0dc30aedde104432cc900f2f0a09eb4fb35cc3}{0x14d...cc3} & 7,660 & 13,330 & 33 & 96 \\

& & Attack-Related & Request after security breaches, e.g., fund recovery, bounty offers. & \hlhref{https://etherscan.io/tx/0x91b41d5b5b30a941c65d5da952c4b11390f364b41cf81a1a87fba9b67b69fe22}{0x91b...e22} & 3,067 & 5,815 & 131 & 174 \\
 
& & Public Apology & Apologies issued by protocols or teams in response to incidents. & \hlhref{https://etherscan.io/tx/0x0efc78a87b4176ee7f2f561e18edd0ec45fd4de1632ccdfca58f8590d602e473}{0x0ef...473} & 119 & 135 & 2 & 3 \\
 
\midrule

\multirow{5}{*}{\makecell{\textbf{Promotion \& }\\ \textbf{Marketing}}}  & \multirow{5}{*}{\cellbar{15.7}} & Project Promotion & Promotion links to protocol websites, token sales, or features. & \hlhref{https://etherscan.io/tx/0xa1de2c1efd6ccf36221cc33bb7eeb29443d24556fdb42bc538612272d3f3eafe}{0xa1d...afe} & 6,316 & 56,367 & 108 & 175 \\

& & Community Invite & Invitations to join social communities, e.g., Telegram, Discord. & \hlhref{https://etherscan.io/tx/0xe11184593f0d869a73f41731e9c636a8d4e63de237657ce3ad9f22ba1a784b70}{0xe11...b70} & 1,432 & 3,371 & 21 & 54 \\
 
& & Token Hype & Hype-building messages about specific tokens or NFTs. & \hlhref{https://etherscan.io/tx/0xe884d4920fb4227cde7ba8565b26718287692d1bcd91f16c6d18bb841ebfd943}{0xe88...943} & 1,052 & 1,663 & 17 & 18 \\
 
& & Airdrop Promotion & Messages promoting token airdrops, e.g., eligibility instructions. & \hlhref{https://etherscan.io/tx/0xeb9f5a02643735b7a56df573821c9490861bf0b9021cd2b332716d8c692310b4}{0xeb9...0b4} & 248 & 2,205 & 8 & 25 \\
 
& & Referral Campaign & Referral campaigns by codes, incentives, or multi-level rewards. & \hlhref{https://etherscan.io/tx/0x44b7239e5fe545a77b4f4836fddaf5fd4a30074df72669b8d47b26309d6b76f4}{0x44b...6f4} & 55 & 3,734 & 2 & 2 \\
 
\midrule

\multirow{3}{*}{\makecell{\textbf{On-chain} \\ \textbf{Requests}}} & \multirow{3}{*}{\cellbar{9.6}} & Fund-related Request & Request for mistaken transfers, lost funds, financial assistance. & \hlhref{https://etherscan.io/tx/0xc6de054b3e8b04084de8cbd7bf515a9a2a180f3f97e40b73bc0ffcb56804c48e}{0xc6d...48e} & 4,766 & 184,378 & 260 & 327 \\

& & Technical Help Request & Requests for technical assistance. & \hlhref{https://etherscan.io/tx/0xe1ead92875e47764ffbd177cb0b5cdcf1fa09b5d0428d513b55be35292e79755}{0xe1e...755} & 346 & 520 & 3 & 3 \\
 
& & Social Support Request & Seeking moral or social support. & \hlhref{https://etherscan.io/tx/0xf98e1e833a0675beec0b6edd31f89733f12a0221d872e8e6f00e46ad79cae008}{0xf98...008} & 269 & 369 & 13 & 13 \\
 
\midrule

\multirow{4}{*}{\makecell{\textbf{Spam/} \\ \textbf{Obfuscation}}} & \multirow{4}{*}{\cellbar{7.6}}  &  Ambiguous Content & Syntactically readable but semantically contextless. & \hlhref{https://etherscan.io/tx/0x07e2621b4d2e2e60848fd47792c9c231d2e4686a1cb8bda919d76b8bf4d4c977}{0x07e...977} & 3,207 & 93,143 & 560 & 790 \\

& & Unreadable Content & Encoded data in formats such as hexadecimal, base64, or binary. & \hlhref{https://etherscan.io/tx/0x2b0275d1190d0cc7844d25ade0415d2e117cddcdcca6ab9a03ff5b0f45ac0763}{0x2b0...763} & 47 & 50 & 480 & 485 \\
 
& & Garbage Content & Sequences of meaningless characters, lacking linguistic structure. & \hlhref{https://etherscan.io/tx/0xd53ca0729a83822170a882cc1ae80690e0bce8a55eb766b4def8cca0396d6b7a}{0xd53...b7a} & 111 & 227 & 71 & 86 \\
 
& & Emoji Flood & Messages composed primarily or entirely of emoji characters. & \hlhref{https://etherscan.io/tx/0x5c0575b185a8b3e1bdb1b718a21de02858d44bcf7b931de5499aecc5f0a9c890}{0x5c0...890} & 11 & 11 & 6 & 6 \\
\midrule

\begin{tabular}{c}
\textbf{On-chain} \\ 
\textbf{Certificate}
\end{tabular}  
& \cellbar{5.5}  & Copyright Certificate & Claims of authorship or copyright made on-chain. & \hlhref{https://etherscan.io/tx/0x5aaa16f6e3e4253f0e3935c61dfd57d8d16ccbbc81fda8f950d8d9b726aaa612}{0x5aa...612} & 0 & 0 & 3,256 & 3,256 \\

\midrule

\multirow{5}{*}{\makecell{\textbf{Financial}\\ \textbf{Content}}} & \multirow{5}{*}{\cellbar{4.8}} & Financial Activity & Expressions of intent to buy, sell, claim tokens, or seek liquidity. & \hlhref{https://etherscan.io/tx/0xcf4f58faf47847d5c86d09f119eef60f642b3a381913fa4c38b0ab7f956b6b35}{0xcf4...b35} & 1,898 & 15,980 & 223 & 490 \\

& & Financial Asset & Descriptions or identifiers of tokens, NFTs, or other assets. & \hlhref{https://etherscan.io/tx/0x680281e7514fb7dc2eaf76d07c2f303d1b61e3af9111863038f8c34b72a3a90a}{0x680...90a} & 0 & 0 & 461 & 490 \\
 
& & Financial Data & Content containing numerical or factual financial data. & \hlhref{https://etherscan.io/tx/0x2b74868463afe2af11fef19b14680fa18a29fd7935eda477f5b90a7c39452cf3}{0x2b7...cf3} & 95 & 112 & 0 & 0 \\
 
& & Financial Analysis & Analytical comments on financial trends, prices, or markets. & \hlhref{https://etherscan.io/tx/0x88d0993639901d61caf3928f73af7b7ced214cd7499b949705f9889afbb9f9ae}{0x88d...9ae} & 34 & 39 & 57 & 67 \\
 
& & Financial Transaction & Messages recording specific financial transactions. & \hlhref{https://etherscan.io/tx/0x662be2562ad08b23e49982422cace93719842522d14f960719c940405bd70354}{0x662...354} & 28 & 28 & 22 & 22 \\
 
\midrule

\multirow{5}{*}{\makecell{\textbf{Toxic/}\\ \textbf{Abusive}\\\textbf{Content}}} &\multirow{5}{*}{\cellbar{4.3}}    & Verbal Abuse \& Profanity  & Offensive or vulgar language intended to insult or provoke. & {0xd9f...f83} & 1,961 & 2,582 & 97 & 103 \\

& & Threats, Harassment \& Psych & Language to threaten, harass, or cause psychological distress. & {0xa9a...8e7}  & 215 & 268 & 116 & 116 \\
 
& & Discriminatory/Hate Speech & Prejudiced language targeting race, gender, religion, or more. & {0x6f3...4b2} & 68 & 77 & 2 & 2 \\
 
& & Sexual/Pornographic Content & Sexual content, including links or solicitation. & {0xbb6...07c} & 36 & 50 & 32 & 32 \\
 
& & Hidden Channel Links & Obfuscated links for scams, adult content, or phishing. & {0xb87...294} & 13 & 126 & 0 & 0 \\
 
\midrule

\multirow{5}{*}{\makecell{\textbf{Cultural/} \\\textbf{Political}\\  \textbf{Expression}}} &  \multirow{5}{*}{\cellbar{4.0}} & Ideological Messaging & Statements of liberty or decentralization. & \hlhref{https://etherscan.io/tx/0x2650c318834c74b9b5c995367bf227c96733992abe293dbfceebf790833e4f4b}{0x265...f4b} & 1,111 & 1,321 & 166 & 319 \\

& & Geopolitical Statement & Takes on wars, governments, or global issues. & \hlhref{https://etherscan.io/tx/0x3ef83021f14516ee1ab4a9aead997fa7d53e524366761870bab6c90937c9bdd9}{0x3ef...dd9} & 280 & 326 & 329 & 488 \\
 
& & Religious Expression & Statements of faith, blessings, religious greetings. & \hlhref{https://etherscan.io/tx/0x67226a6688f419d5a1033aeefb4b579a92e16a91a98a14bf37ad71998f2e3451}{0x672...451} &  351 & 484 & 58 & 66 \\
 
& & Political Slogans & Protest phrases or activism slogans. & \hlhref{https://etherscan.io/tx/0x53bff0c77c555283ded3caa849fcd335553f4039181f3153663b003f5c175a72}{0x53b...a72}  & 28 & 29 & 3 & 3 \\
 
& & Cultural Commentary & Expressing opinions on cultural topics, trends, or values. & \hlhref{https://etherscan.io/tx/0x826383896239eb4f68012a88f149c80f964df1694cc238405caa0a89b00583b0}{0x826...3b0} & 9 & 9 & 11 & 12 \\
 
\midrule

\multirow{4}{*}{\makecell{\textbf{Technical/}\\\textbf{Developer}\\ \textbf{Message}}} &  \multirow{4}{*}{\cellbar{2.9}}  & On-chain Records & Documenting transactions, events, state changes, or gas usage. & \hlhref{https://etherscan.io/tx/0x6a43512e75d34cbd83cd4e0bd3c355dc497da1b1e90275b3ad66b752fac528e6}{0x6a4...8e6} & 981 & 2,342 & 71 & 89 \\

& & Deployment Notice & Announcements related to smart contract deployment. & \hlhref{https://etherscan.io/tx/0xc1be9ba5a6b6f81298a4fdadca3fa2f37797ce580b21a592a418349b05acccc5}{0xc1b...cc5} & 275 & 463 & 7 & 7 \\
 
& & Testing Notice & Messages indicating testing activity or test data injection. & \hlhref{https://etherscan.io/tx/0x90cc235d442b9be16170fd84b324dd9b503c461a0b445b38b2d0152e8c4a0b81}{0x90c...b81} & 195 & 349 & 20 & 22 \\
 
& & Code Snippets & Embedded fragments of source code, settings, or function logic. & \hlhref{https://etherscan.io/tx/0x8fd78ff78b1cd59a4449224fb29d6d80b69555d8b91f756559cb82df7cbc7e58}{0x8fd...e58} & 131 & 145 & 4 & 4 \\
 
\midrule

\multirow{2}{*}{\makecell{\textbf{Education}}} &  \multirow{2}{*}{\cellbar{1.1}}  & Course Completion & Declarations or proofs of course or degree completion. & \hlhref{https://etherscan.io/tx/0x067d08658bc3b0fd8aeffc13eee9010c840658a633ba38d68115896cebe5d96c}{0x067...96c} & 241 & 250 & 293 & 305 \\

 & & Knowledge Sharing & Technical knowledge, explanations, or community support. & \hlhref{https://etherscan.io/tx/0xb15cd8e2d1efe442638a047095d7fda5eeaf310a6b1eb066012a50a36e27a6fd}{0xb15...6fd} & 59 & 62 & 75 & 91 \\
 
\midrule

\multirow{2}{*}{\makecell{\textbf{Charity/}\\\textbf{Fundraising}}}  & \multirow{2}{*}{\cellbar{0.3}} & Help Request & User-initiated appeals for fundraising, or charitable support. & \hlhref{https://etherscan.io/tx/0xe8fe9d776e323eb27b1eaba371363d9b52ad9c2162f29852897107e95d6813f7}{0xe8f...3f7} & 119 & 269 & 8 & 8 \\

& & Records & Messages recording donations made or funds received. & \hlhref{https://etherscan.io/tx/0x6c08e97062fa260f2ced9d10c3a194c92d91d6c3b37060272f2b74b54ab16468}{0x6c0...468} & 38 & 39 & 29 & 42 \\
 
\midrule

\multirow{1}{*}{\textbf{Others}}  & \cellbar{3.1} & Others & Categories not defined & \hlhref{https://etherscan.io/tx/0xb73b7097a33bd96b8198031658b81a4238a1e388a2b10730caac864aff3031ca}{0xb73...1ca} & 1,230 & 1,377 & 597 & 710 \\

\bottomrule
\end{tabular}}
\begin{tablenotes}[flushleft]
      \footnotesize
      \item[] \quad \textit{
      IDM links under the main topic of Toxic/Abusive Content were removed to comply with the Anti-Harassment Policy.
      }
\end{tablenotes}
\end{table*}

\section{IDM Semantic Analysis}
\label{sec: semantic}

This section presents a semantic analysis of \IDMs, focusing on the subset containing English and Chinese natural language. We examine their topical composition and sentiment characteristics.

\subsection{Topic Analysis}
\label{sec: topic_analysis}

\subsubsection{\underline{Methods}}
Given the domain-specific and noisy nature of on-chain messages, we found that unsupervised topic modeling methods were insufficient to generate coherent and meaningful topics. Therefore, we applied a semi-supervised, taxonomy-driven approach to topic classification. A predefined set of main topics and subtopics was constructed based on domain knowledge and the initial exploration of the dataset.  We then utilized a large language model, \texttt{GPT-4o}, to map each \IDM to the most suitable main-subtopic pair, or to propose a new topic pair when necessary. Human review and iterative prompting adjustments were also employed to refine model outputs and ensure consistent labeling quality. This human-in-the-loop strategy ensured both consistency with domain expertise and adaptability to emerging topics.

\begin{figure}[h]
\centering
\includegraphics[width=0.98\columnwidth]{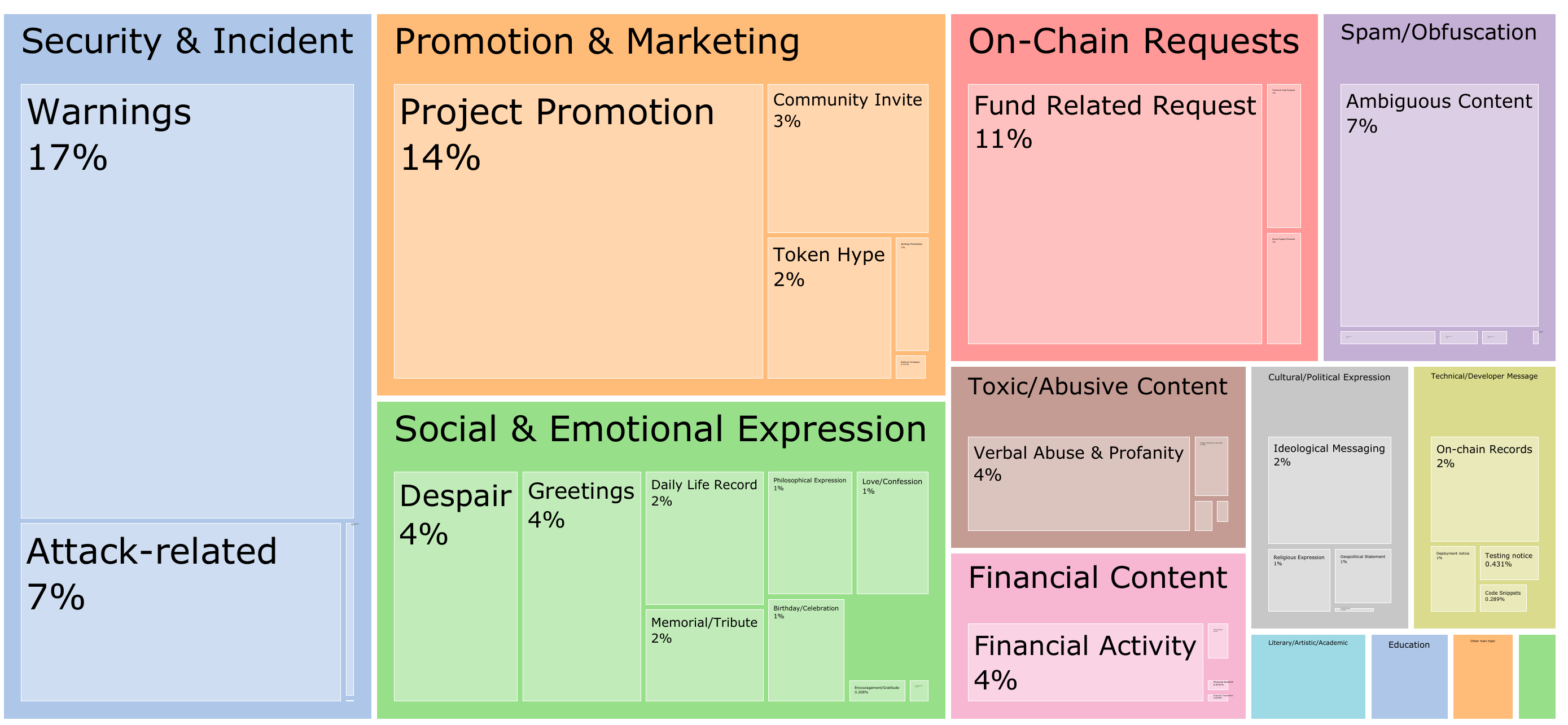}
\caption{Topic distribution for unique English IDMs.}
\label{fig: topic_en}
\end{figure}

\begin{figure}[h]
\centering
\includegraphics[width=0.98\columnwidth]{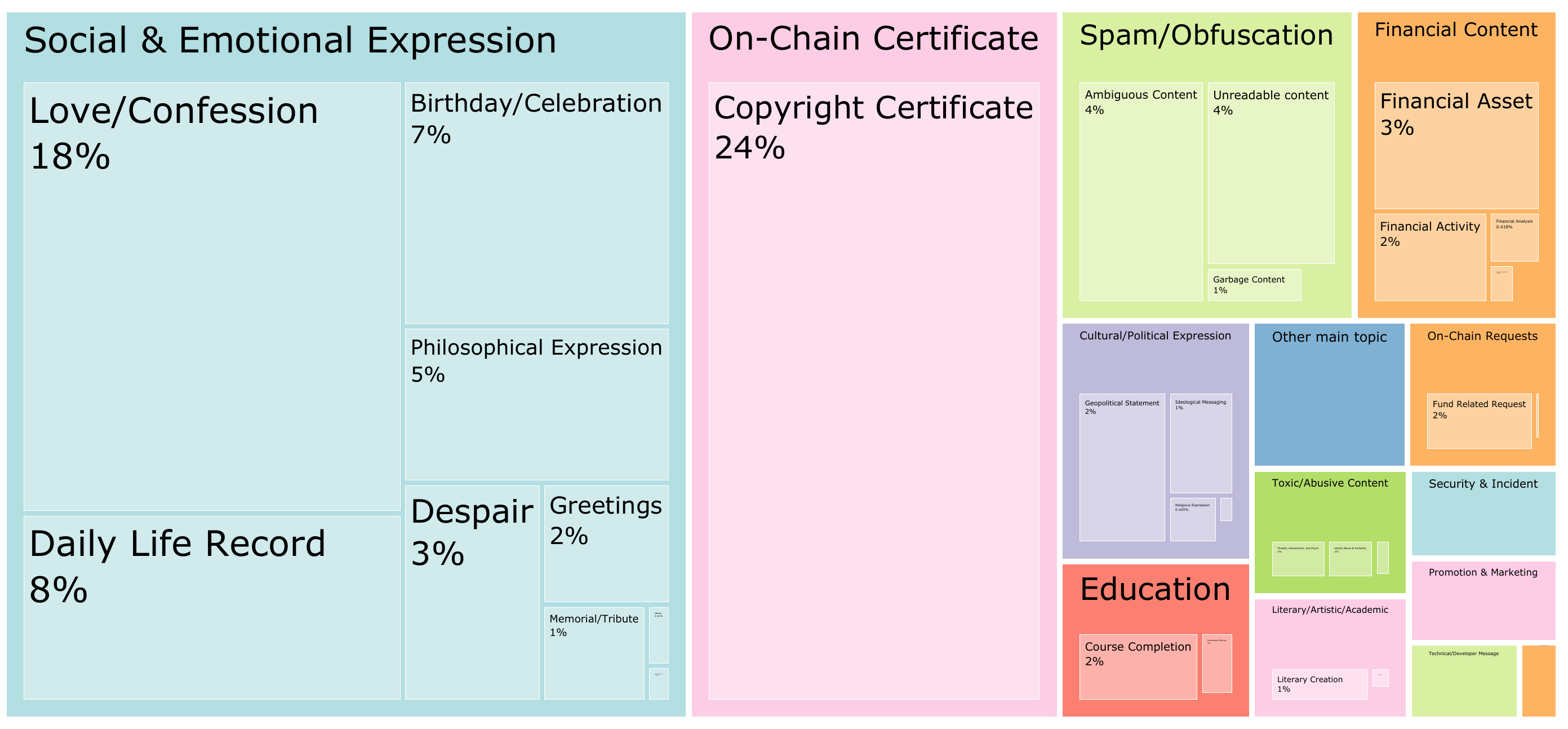}
\caption{Topic distribution for unique Chinese IDMs.}
\label{fig: topic_cn}
\end{figure}

\subsubsection{\underline{Results}} 
Table~\ref{tab:topic_taxonomy} presents the Ethereum \IDM topic taxonomy, organized by main topics and subtopics. 
For classification, we used \texttt{GPT-4o} to assign the most relevant topics to each message. 
We report both the number of unique \IDMs and the total number of transactions for English and Chinese messages under each subtopic.

\smallskip
\noindent\textbf{Topic Distribution.} Our topic taxonomy in Table~\ref{tab:topic_taxonomy} covers $12$ main topics with $48$ subtopics. Interestingly, the distributions of English and Chinese \IDMs are quite different. English \IDMs are more concentrated in \textit{Security \& Incidents} (24\%, see Figure~\ref{fig: topic_en}), whereas Chinese \IDMs are more prevalent in \textit{Social \& Emotional Expression} (44\%, see Figure~\ref{fig: topic_cn}). 
Specifically, English \IDM senders are more likely to use messages to issue warnings about token scams, rug pulls, and honeypots (17\%), promote projects and token sales (14\%), and request assistance related to lost funds or financial hardship (11\%). In contrast, Chinese \IDM senders more often use messages for social connection and emotional expression (44\%).  In fact, 18\% of IDMs express love and affection, 8\% document the sender’s daily life, and 7\% convey birthday wishes or holiday greetings. In addition, 24\% of the Chinese IDMs are used to certify copyright claims related to real-world assets (24\%). This shows that users from different language communities use the Ethereum transaction layer not only for P2P transfers but also for distinct communicative purposes. The \IDM channel functions as both a transactional tool and a means of expression shaped by local norms and intentions.

\begin{figure}[tbh]
\centering
\includegraphics[width=\columnwidth]{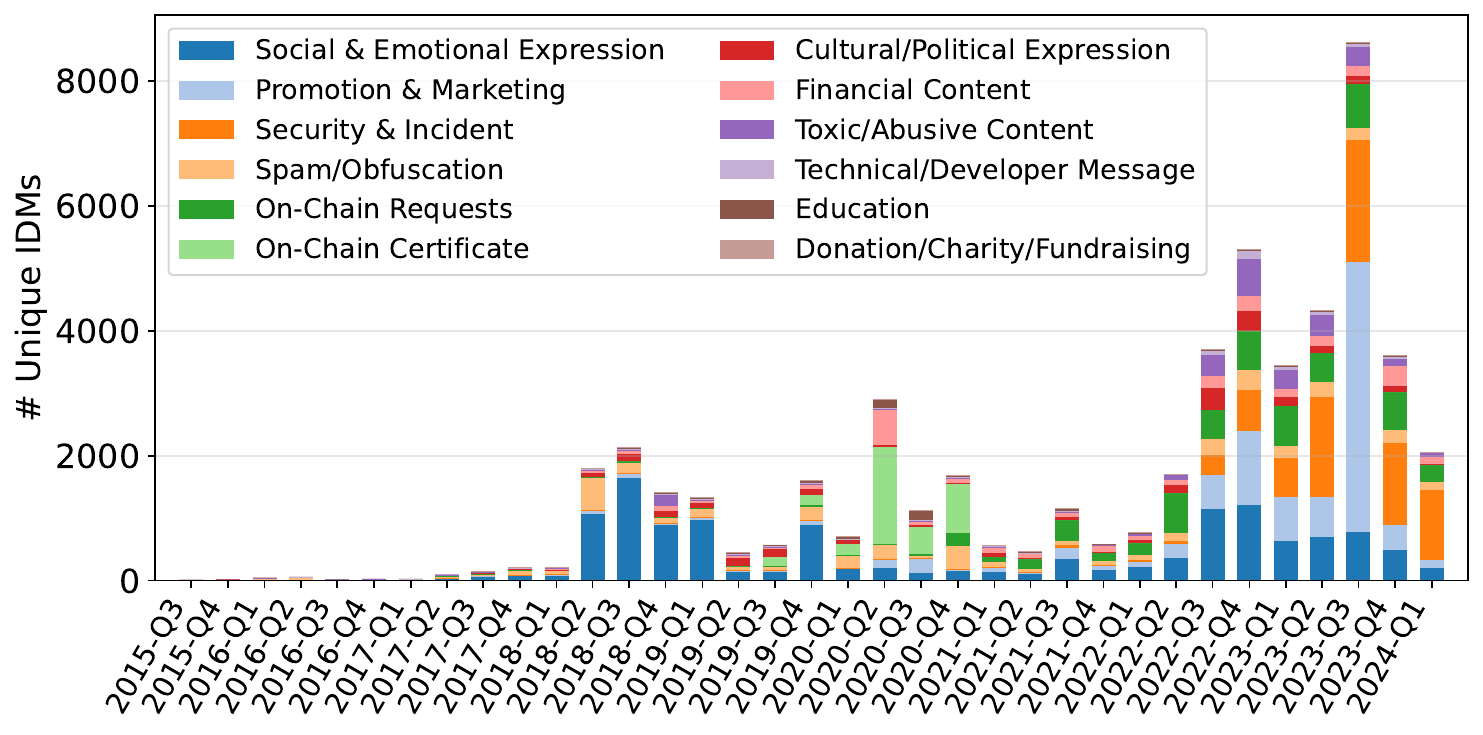}
\caption{Temporal trends of topics (unique IDMs).}
\label{fig:topics_evolution_quarterly}
\end{figure}

\smallskip
\noindent\textbf{Temporal Topic Dynamics.} Figure~\ref{fig:topics_evolution_quarterly} illustrates the distribution of unique \IDMs by semantic topic, aggregated quarterly from 2015 to 2024. The first round of surge occurred in 2018–2019, with \textit{Social \& Emotional Expression} and \textit{Spam/Obfuscation} as the dominant topics. This period likely reflects the experimental and symbolic use of input data, including emotional expression, greetings, and low-quality or repetitive messages. In 2020–2021, there is a shift toward \textit{On-chain Certificates}, indicating the emergence of utility-driven use cases, such as copyright or asset attestation.

Starting in late 2022, a second, broader surge in \IDM volume occurs. This phase is characterized by a rise in \textit{Promotion \& Marketing}, \textit{Security \& Incidents}, and \textit{On-chain Requests}. This reflects a growing tendency to use Ethereum not only for personal or expressive purposes, but also for transactional communication related to community campaigns, security risks, and fund recovery.

The peak in Q3 2023 stands out as the highest observed to date, driven primarily by security and promotion topics. This suggests an intensified use of \IDMs for public messaging. This trend signals a transformation in the communicative role of input data: from informal symbolic use to a more functional, utility-oriented medium.

\subsubsection{\underline{Case Study}} In the following, we provide concrete examples to show that IDM topics are closely related to important social events. These examples suggest that Ethereum can be a medium for social activism and historical documentation.

\smallskip
\noindent\textbf{The \#MeToo Movement in 2018.} 
The global \#MeToo Movement~\cite{rodino2018me} is a social movement against sexual harassment. It encourages survivors to share their experiences publicly and gained widespread momentum in 2018, especially in China. As shown in Figure~\ref{fig:topics_evolution_quarterly}, during Q4 2018, we observe a notable increase in IDMs associated with the topic of \textit{Social \& Emotional Expression}. Through manual investigation, we identify more than $800$ IDMs explicitly linked to the expression of \#MeToo-related messages and slogans.

A representative example can be found in the Chinese IDMs sent out by the Ethereum address {0xd44...857}\footnote{Address link is removed to comply with the Anti-Harrassment Policy.}. These messages include personal disclosure of trauma expression from individuals who have experienced sexual abuse. Interestingly, all of these IDMs were issued within a period of less than ten days, but they include the stories of hundreds of people. This pattern indicates that the IDM senders tend to leverage Ethereum IDMs as a means to record these stories. This allows the voices of \#MeToo survivors to be preserved without the risk of removal by centralized entities.

\smallskip
\noindent\textbf{COVID-19 during 2020 and 2023.}
We identify $149$ IDMs containing COVID-19 words such as ``Covid'' and ``lockdown'' in our dataset. Specifically, these IDMs are distributed across the years 2020 to 2023, with $78$ \IDMs in 2020, $27$ in 2021, $16$ in 2022, and $28$ in 2023. These IDMs contain a range of public reactions and behaviors during the pandemic and mainly include three topics: \emph{(i)} \emph{Cultural and Political Expression}, where individuals voice opinions on public policies and societal responses to COVID-19; \emph{(ii)} \emph{Social and Emotional Expression}, reflecting personal experiences, mental health status, and feelings during COVID-19 lockdowns; and \emph{(iii)} \emph{On-Chain Requests}, in which users leverage Ethereum IDMs to ask for help, share needs, or call for collective action.

\subsection{Sentiment Analysis}
\label{sec: sentiment_analysis}

We developed a taxonomy~(see Table~\ref{tab:emotion_taxonomy}) informed by the \href{https://saifmohammad.com/WebPages/NRC-Emotion-Lexicon.htm}{NRC Emotion Lexicon} and refined it to better fit the context of blockchain communication. The taxonomy includes three polarity classes (\textit{Positive}, \textit{Negative}, \textit{Neutral}), with six positive emotions (e.g., \textit{Joy}), six negative emotions (e.g., \textit{Anger}), and four neutral states (e.g., confusion). This structure allowed for more reliable classification of emotional tone within the communicative patterns found in \IDMs.

\begin{table}[tbh]
\centering
\caption{IDM sentiment taxonomy with polarity, emotion, example IDM, and language-specific counts.}
\label{tab:emotion_taxonomy}
\resizebox{\columnwidth}{!}{
\begin{tabular}{cll|cc|cc}
\toprule
\multirow{2}{*}{\textbf{Polarity}} & \multicolumn{1}{c}{\multirow{2}{*}{\textbf{Emotion}}} & \multirow{2}{*}{\textbf{Example}} & \multicolumn{2}{c|}{\textbf{English}} & \multicolumn{2}{c}{\textbf{Chinese}} \\
\cmidrule(lr){4-5} \cmidrule(lr){6-7}
& & & \# Unique & \# Total & \# Unique & \# Total \\
\midrule
\multirow{6}{*}{Positive} & Joy & \href{https://etherscan.io/tx/0x8a65ed397a9b25cf9a4bfcb487fb68c0bd4027527a56b899ab9a46d525d13182}{0x8a6...182} & \cellcolor[HTML]{FDCFA0}1{,}831 & \cellcolor[HTML]{DDEAF7}9{,}276 & \cellcolor[HTML]{FD974A}4{,}220 & \cellcolor[HTML]{E0ECF8}4{,}335 \\
& Trust & \href{https://etherscan.io/tx/0xd7683c8392113f9828e9500b8d5a7374a7b857e139af1855d25af7db5736aa47}{0xd76...a47}  & \cellcolor[HTML]{FDCE9E}1{,}880 & \cellcolor[HTML]{E0ECF8}4{,}140 & \cellcolor[HTML]{FEE7D1}215 & \cellcolor[HTML]{E3EEF9}318 \\
& Love & \href{https://etherscan.io/tx/0x98ed25c1b78b80f787fb5d4c6917962745fd1cbade866adf619632d5982924ae}{0x98e...4ae} & \cellcolor[HTML]{FEE0C1}754 & \cellcolor[HTML]{E3EEF8}907 & \cellcolor[HTML]{FDBA7F}2{,}681 & \cellcolor[HTML]{DFECF7}5{,}553 \\
& Hope & \href{https://etherscan.io/tx/0x8f59d739900f35fb7cdfd65e6fd04dccebd13cc4feeed72f13ecb6b564cfe3f7}{0x8f5...3f7} & \cellcolor[HTML]{FD9F56}3{,}879 & \cellcolor[HTML]{DCE9F6}11{,}175 & \cellcolor[HTML]{FEDCB9}1{,}021 & \cellcolor[HTML]{E3EEF8}1{,}295 \\
 & Gratitude & \href{https://etherscan.io/tx/0xce7ef51c8aa2424cd34d782b20688d5eb32dcc7786f2064df129e29a81973e1c}{0xce7...e1c}  & \cellcolor[HTML]{FDD5AB}1{,}475 & \cellcolor[HTML]{E1EDF8}3{,}188 & \cellcolor[HTML]{FEE7D0}256 & \cellcolor[HTML]{E3EEF9}299 \\
 & Warmth & \href{https://etherscan.io/tx/0x9b2522744cca40bef55cb1813dc6f96128eb6da0880fb31af8d3e15eb16f178e}{0x9b2...78e}  & \cellcolor[HTML]{FDD7AF}1{,}349 & \cellcolor[HTML]{E1EDF8}3{,}695 & \cellcolor[HTML]{FEE0C3}698 & \cellcolor[HTML]{E3EEF8}786 \\
\midrule
\multirow{6}{*}{Negative} & Anger &\href{https://etherscan.io/tx/0x637e7ee0e60b2fba8531cd8846fcdc2cea5a47ac5daa7b6307f98fda03270cb3}{0x637...cb3}  & \cellcolor[HTML]{FDD9B4}1{,}209 & \cellcolor[HTML]{E3EEF8}1{,}628 & \cellcolor[HTML]{FEE8D2}119 & \cellcolor[HTML]{E3EEF9}132 \\
 & Hostility & {0xcd6...554} & \cellcolor[HTML]{FDBA7F}2{,}700 & \cellcolor[HTML]{E1EDF8}3{,}952 & \cellcolor[HTML]{FEE8D2}93 & \cellcolor[HTML]{E3EEF9}105 \\
 & Disgust & {0xf52...f83} & \cellcolor[HTML]{FDD8B2}1{,}258 & \cellcolor[HTML]{E2EDF8}1{,}814 & \cellcolor[HTML]{FEE7D0}240 & \cellcolor[HTML]{E3EEF9}264 \\
 & Fear & \href{https://etherscan.io/tx/0xba12cbdc61d8ad177d939fc61213771cfb81b07e9202bc47aa51d8fc36e16e4d}{0xba1...e4d}  & \cellcolor{red!60}7{,}558 & \cellcolor{cyan!40}196{,}338 & \cellcolor[HTML]{FEE6CE}332 & \cellcolor[HTML]{E3EEF9}414 \\
 & Guilty & \href{https://etherscan.io/tx/0x80185b9aac223506c414c77914a71f9e27fa7c53c345e92dfc4128fb2187e6a1}{0x801...6a1}  & \cellcolor[HTML]{FEE5CB}419 & \cellcolor[HTML]{E3EEF8}521 & \cellcolor[HTML]{FEE9D4}31 & \cellcolor[HTML]{E3EEF9}31 \\
 & Sadness & \href{https://etherscan.io/tx/0x64a84b7b4ec87aebb45b078939c5480be4736ebb86bfe3cf3130e499840d5e53}{0x64a...e53}  & \cellcolor[HTML]{FDB373}3{,}014 & \cellcolor[HTML]{E0ECF8}4{,}392 & \cellcolor[HTML]{FEDFC0}771 & \cellcolor[HTML]{E3EEF8}894 \\
\midrule
\multirow{4}{*}{Neutral} & Surprise & \href{https://etherscan.io/tx/0x39d9d5a47870f469efff0bee54452e9d516cc15831aaab50f7f837ecf2353d46}{0x39d...d46}  & \cellcolor[HTML]{FEE7D0}265 & \cellcolor[HTML]{E3EEF9}362 & \cellcolor[HTML]{FEE9D4}26 & \cellcolor[HTML]{E3EEF9}28 \\
& Confusion & \href{https://etherscan.io/tx/0x22369b269595dcb20ab71ceecf534622457b96339d516fa36837eba3e3335da3}{0x223...da3} & \cellcolor[HTML]{FEE0C3}723 & \cellcolor[HTML]{E3EEF8}1{,}273 & \cellcolor[HTML]{FEE9D3}69 & \cellcolor[HTML]{E3EEF9}97 \\
& Curiosity & \href{https://etherscan.io/tx/0xc7f3c64b8bfffb96182c8726f5fd9bf4b401ad623b5912555ee7a802064beaf0}{0xc7f...af0}  & \cellcolor[HTML]{FDA965}3{,}400 & \cellcolor[HTML]{DFECF7}5{,}290 & \cellcolor[HTML]{FEE5CB}414 & \cellcolor[HTML]{E3EEF8}648 \\
& Politeness & \href{https://etherscan.io/tx/0xc3a5a177686a0c59b65bc797cac862b6f348048bb0879e50301a08ae4d9c0598}{0xc3a...598} & \cellcolor[HTML]{FDA863}3{,}483 & \cellcolor[HTML]{D6E5F4}21{,}494 & \cellcolor[HTML]{FEE8D2}154 & \cellcolor[HTML]{E3EEF9}171 \\
\bottomrule
\end{tabular}
}
\begin{tablenotes}[flushleft]
      \footnotesize
      \item[] \quad \textit{
      Color intensity indicates count magnitude. For Hostility and Disgust, IDM links are removed to comply with the Anti-Harassment Policy.
      }
\end{tablenotes}
\end{table}

For sentiment analysis, traditional tools such as lexicon-based methods or pretrained classifiers often struggle with domain-specific, noisy text. In the case of IDMs, these tools are limited due to the presence of on-chain identifiers, irregular formatting, and mixed linguistic cues. To address these challenges, we used OpenAI's \texttt{GPT-4o} model to perform emotion classification. For each \IDM, the model was prompted to return the relevant emotion with the associated intensity score (1–10) and confidence value (0–1).

We have identified \numENIDMwithEmotion emotive \IDMs in English and \numCNIDMwithEmotion emotive \IDMs in Chinese. Table~\ref{tab:emotion_taxonomy}  presents the emotion distribution of English and Chinese \IDMs, categorized by polarity and specific emotion types. Negative emotions are more prevalent in English \IDMs, whereas Chinese \IDMs show significantly fewer instances of such expressions. This observation indicates a higher tendency to express negative sentiment in English \IDMs.

\textit{Fear} stands out as the most prevalent emotion category for English \IDMs, with $7{,}558$ unique messages and $196{,}338$ total occurrences, far surpassing all other emotional labels. This pattern reflects widespread usage of \IDMs for warnings, scam alerts, or distress signals.  It also underscores persistent concerns over on-chain security, as users frequently respond to phishing, hacks, and contract vulnerabilities with messages conveying fear and urgency. \textit{Hostility} is also prominent in English \IDMs, indicating the potential use of the IDM channel for toxic expression.

In contrast, Chinese \IDMs are more concentrated in positive emotional expressions, particularly \textit{Joy} (4,220 unique) and \textit{Love} (2,681 unique), suggesting a more affective and interpersonal use of on-chain messaging in the Chinese context.

\begin{figure}[tbh]
\centering
\includegraphics[width=\columnwidth]{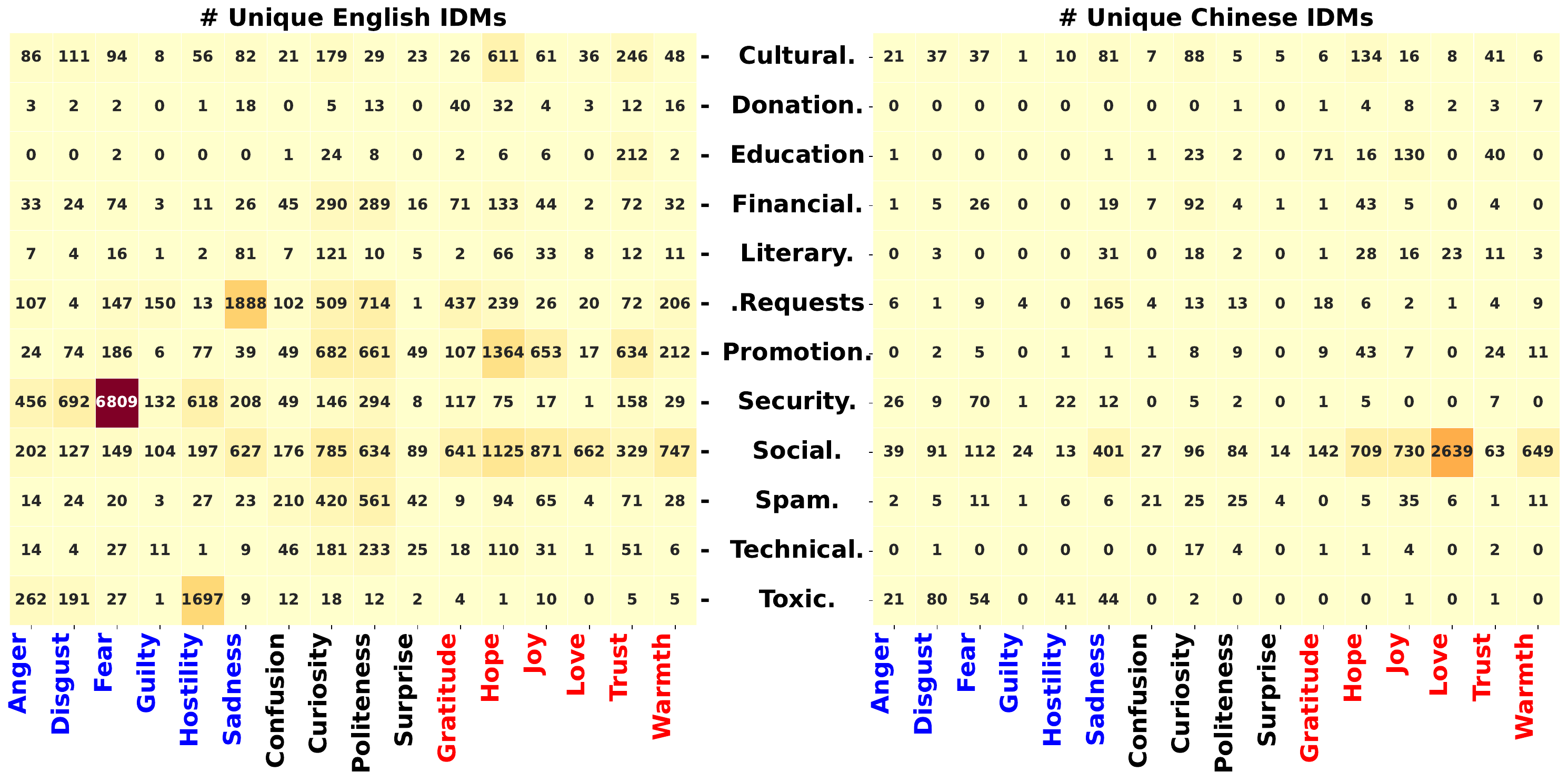}
\caption{Emotion -- topic distribution of unique Ethereum IDMs in English (left) and Chinese (right).}
\label{fig:topic_sentiment_heatmap}
\end{figure}

We further present a cross-tabulation of emotional categories and IDM topics in Figure~\ref{fig:topic_sentiment_heatmap}. 
English IDMs show strong emotional variance across topics.  Notably, \textit{Fear} is dominant in \textit{Security \& Incident} (6,809 messages), suggesting that discussions around scams, hacks, or risk tend to evoke anxiety-driven expressions. \textit{Hostility} dominates within \textit{Toxic/Abusive Content}, suggesting that IDMs are often used to convey verbal aggression or threats. In addition, \textit{Sadness} appears most frequently in \textit{On-chain Requests}, reflecting expressions of personal hardship or appeals for help.

By contrast, Chinese \IDMs show emotional clustering in \textit{Social \& Emotional Expression}, particularly in \textit{Love} (2,639), \textit{Joy} (730), and \textit{Hope} (709).  Negative emotions are far less pronounced across Chinese topics. This reinforces earlier findings that Chinese users more often utilize \IDMs for interpersonal or emotional communication.

Overall, the matrix reveals a clear language-specific divergence. English \IDMs carry heavier negative emotional signals in security-related and toxic content, whereas Chinese \IDMs reflect more interpersonal and affective sentiment within social topics.

\section{IDM Communication Costs}
\label{sec-idmComm}

In this section, we analyze IDM transaction value and cost.

\subsection{Transaction Values}

Different from traditional social media platforms, Ethereum \IDMs are propagated via blockchain transactions, which can typically trigger the transfer of coins such as \ETH. As a result, \IDMs transactions serve not only as a medium for information exchange but also as a mechanism for asset transfer. This dual role, i.e., communicative and transactional, introduces measurable costs that vary with IDM characteristics such as topic, length, and language.

\begin{figure}[htb]
    \centering
    \includegraphics[width=0.95\columnwidth]{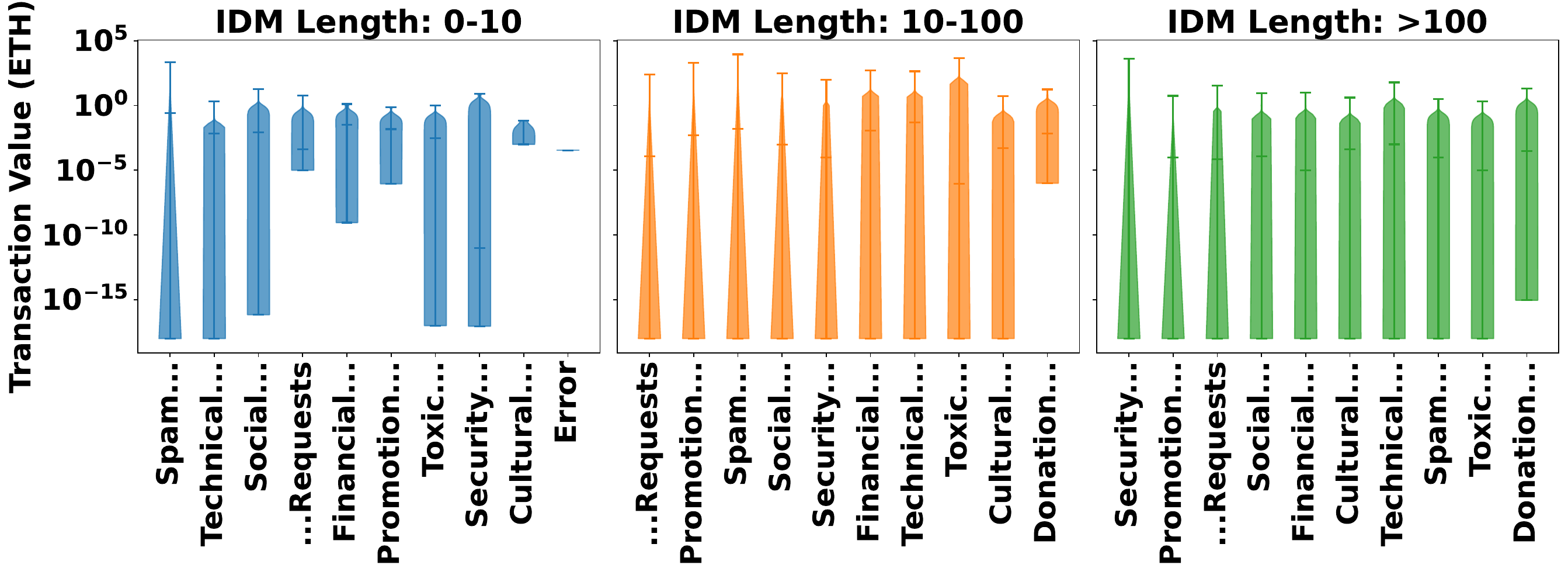}
    \caption{English IDM topic and transaction values.}
    \label{fig:english_topic_value_violin}

    \vspace{0.2em} 

    \includegraphics[width=0.95\columnwidth]{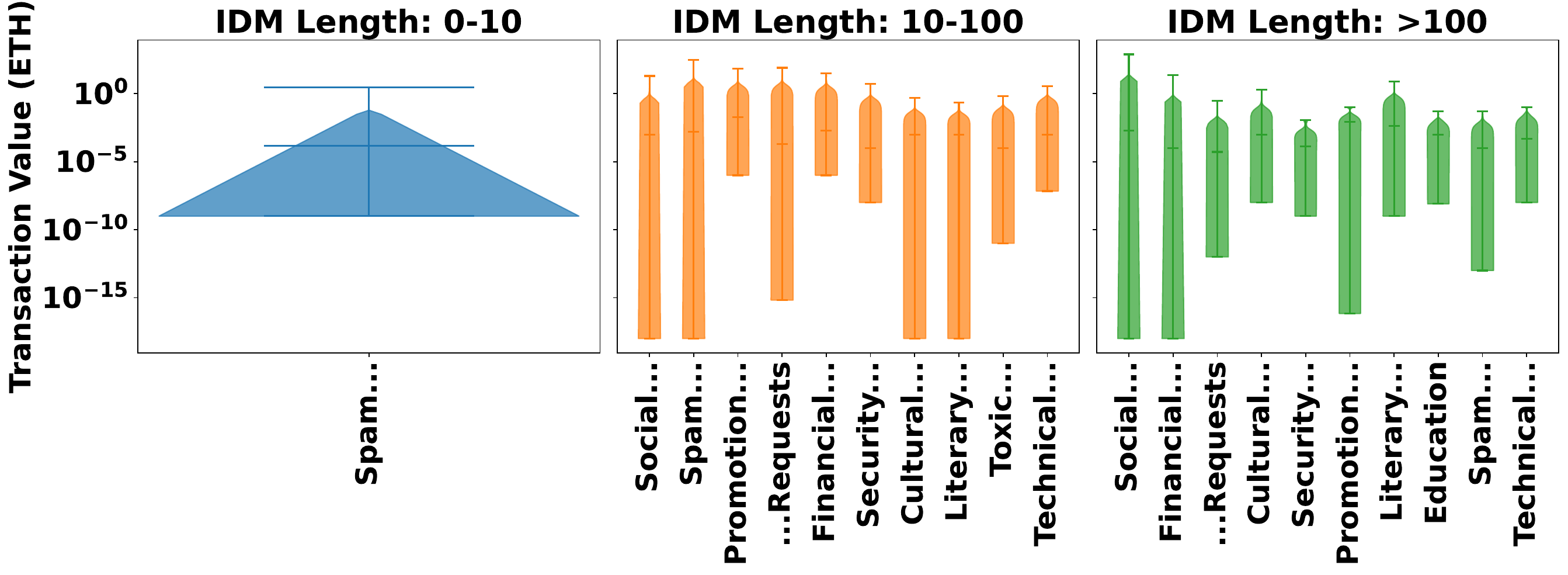}
    \caption{Chinese IDM topic and transaction values.}
    \label{fig:chinese_topic_value_violin}
\end{figure}

In total, we identify $340{,}900$ non-zero-value transactions with English and Chinese \IDMs.  Figures~\ref{fig:english_topic_value_violin}
and~\ref{fig:chinese_topic_value_violin} show the distribution of transaction values and topics for English and Chinese IDMs, segmented by message length (i.e., $0$ – $10$, $10$ – $100$,  $\ge$$100$ bytes). Within each length category,  we select the top $10$ most frequent topics and visualize their distribution over transaction values.

In English IDMs (see Figure~\ref{fig:english_topic_value_violin}), the dominant topics in short messages ($0$ – $10$ bytes) are \textit{Spam/Obfuscation}, \textit{Technical \& Developer Messages}, and \textit{Social \& Emotional Expression}. We also observe that transaction values range widely and include several high-value outliers. As message length increases ($10$ – $100$ bytes), dominant topics shift toward \textit{On-Chain Requests} and \textit{Promotion \& Marketing}. In addition, long-message (> $ 100$ bytes) transactions tend to carry more \ETH transfers compared to those with shorter messages, and the most popular topic now is \textit{Security \& Incident}. For example, the IDM transaction \href{https://etherscan.io/tx/0x370f1f6b5f6136174c01a26c63a7f01eedb99b1b09d44b5f640acb0bcf37daf4}{0x370...af4} with the largest value (i.e., $3{,}997.9$ \ETH) is issued by the HTX Global Hacker, which is used to send back the attack revenue to the Huobi Recovery address.

In Chinese IDMs (see Figure~\ref{fig:chinese_topic_value_violin}), transaction values for short messages are generally negligible, even across common topics such as \textit{Spam/Obfuscation}. Similar to the cases in English IDMs, as the length increases, topics such as \textit{On-Chain Requests}, \textit{Promotion \& Marketing}, and \textit{Security \& Incident} emerge, with moderate transaction values. Notably, among IDMs with a message length of 10 bytes or more, the most common topic is \textit{Social \& Emotional Expression}. This highlights a cultural tendency to use \IDMs for self-expression. For instance, we observe that the \IDMs with specific transaction values of $5.203344$, $5.201314$, $5.20$, $1.314$, $0.5201314$, and $0.1314$ are often used to express love and affection. This follows popular symbolic numerology in contemporary Chinese digital culture\footnote{In Chinese, ``520'' is a homophonic representation of ``wǒ ài nǐ'' (``I love you''), and ``1314/3344'' corresponds to ``yī shēng yī shì/shēng shēng shì shì'' (``a whole lifetime''). }. For example, in transaction \href{https://etherscan.io/tx/0x2889a749cc4ef593486f1fb3c60a1cc21eb1d1b94a7a0972411aa4dc958ede0a}{0x288...e0a}, the sender sends 5.201314 ETH to their spouse to commemorate their wedding anniversary.

The comparison reveals different communicative norms. Long English \IDMs tend to serve transactional or protocol-level purposes, while long Chinese \IDMs convey stronger emotional intent.

\subsection{Transaction Costs}

\begin{figure}[htb]
\centering
\includegraphics[width=\columnwidth]{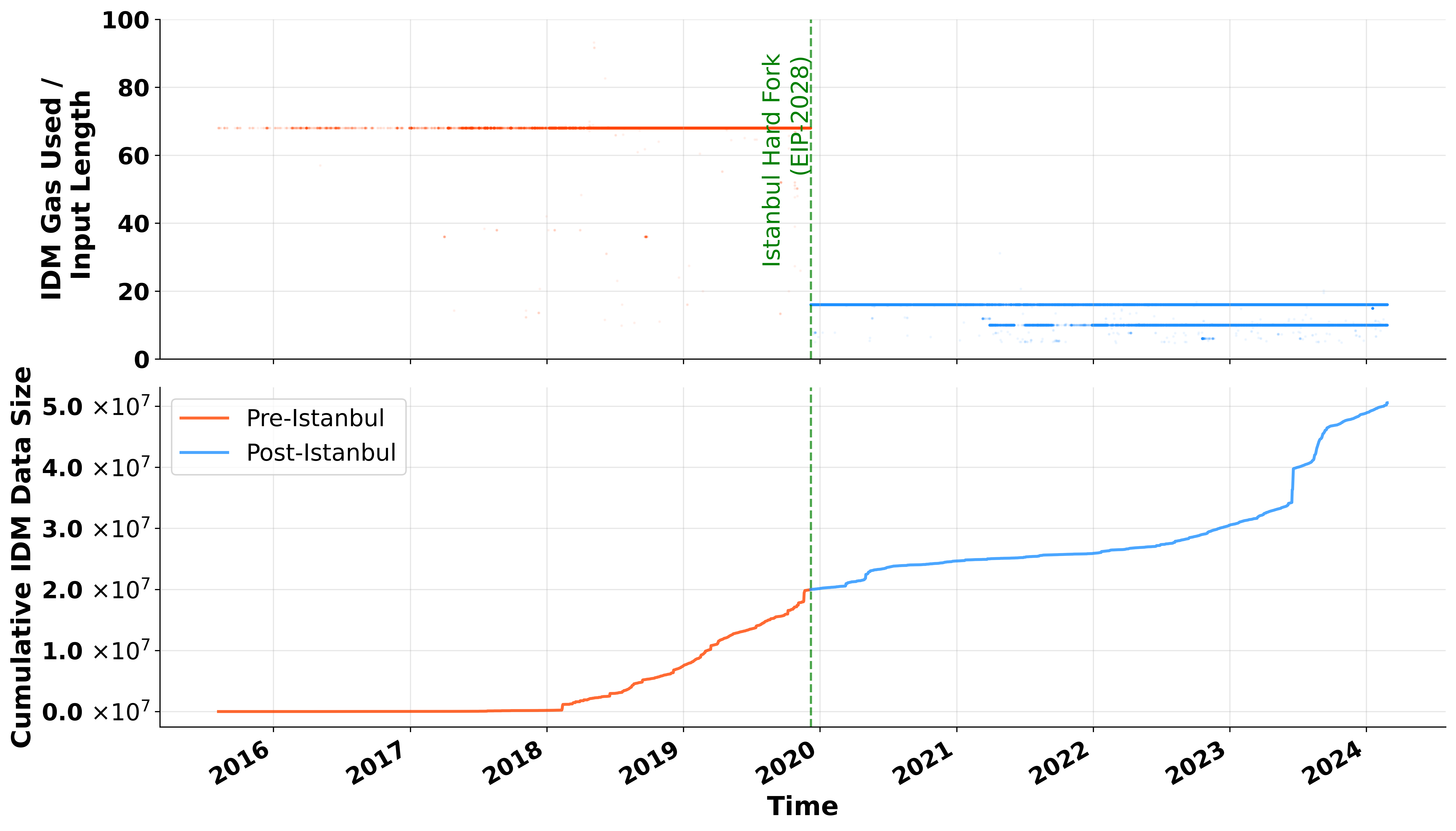}
\caption{IDM gas costs and cumulative data size over time. The differences between the red and the blue lines in the left subfigure are due to the Istanbul upgrade.}
\label{fig:gas_ratio_and_input_length}
\end{figure}

Compared to basic \ETH transfers (which typically require $21{,} 000$ gas), sending an IDM triggers additional gas costs for storing message data on-chain. Figure~\ref{fig:gas_ratio_and_input_length} provides a temporal analysis of gas usage and data size trends for IDM transactions. 

The top subfigure plots the gas used per byte of \IDM input over time. Before the Istanbul hard fork~\cite{istabul2019} in 2019, the per-byte gas cost remained consistently high, which was around $68$ gas units per byte (in the red band). Following the upgrade, this cost dropped sharply to $16$ gas units per byte, resulting in the much lower blue band observed post-2020. This significant reduction indicates how EIP-2028~\cite{eip2028} improved the efficiency of transactions with \IDMs.

The bottom subfigure shows the cumulative size of IDM data over time. Although the Istanbul upgrade in late 2019 reduced the gas cost per byte, its direct impact on the volume of IDM data was limited. The figure shows that a notable acceleration in IDM data accumulation did not occur until mid-2023. We suspect that factors beyond gas efficiency, such as broader adoption of social messaging patterns, might play a more critical role in the increase.

In addition, this increase in IDM data volume raises ongoing debates within the Ethereum community regarding the utility and long-term impact of such content. Unlike smart contract interactions or financial transfers, a large proportion of IDM messages, such as those expressing personal sentiments or opinions, do not directly contribute to the blockchain's core financial functionalities. Nonetheless, these messages are permanently recorded on-chain, which will occupy the storage space on Ethereum nodes.

\section{IDM Network Analysis}
\label{sec-idmNetwork}
The IDM senders and receivers form a communication network. In this section, we analyze the structure of the Ethereum IDM network.

\subsection{Network Measurement}

In total, we identify that \TotalAddrsWithNaturalTextTXs addresses\footnote{Some EOAs send IDMs only to others, some only to themselves, and others to both.} send \TotalWithNaturalTextTXs transactions that contain natural languages, where \UniqueWithNaturalTextTXs IDMs are unique. 

\begin{figure}[htb]
\centering
\includegraphics[width=\columnwidth]{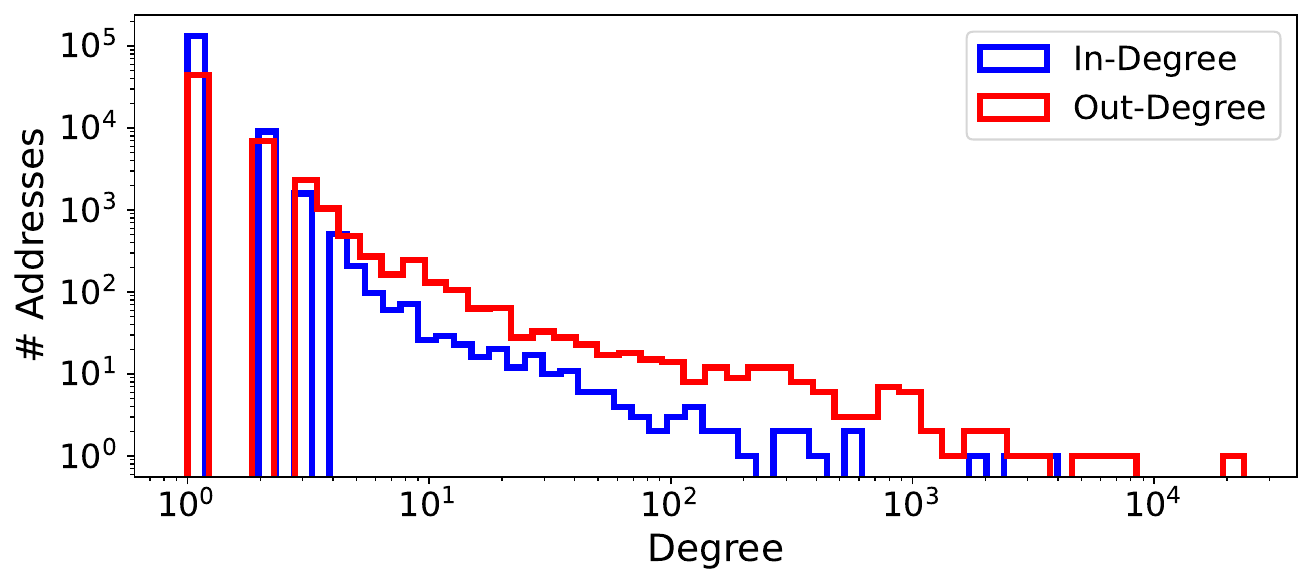}
\caption{Distribution of indegree and outdegree.}
\label{fig:Degree_Distribution}
\end{figure}

\smallskip
\noindent\noindent\textbf{Network Degree Distribution.} 
As shown in Figure~\ref{fig:Degree_Distribution}, the IDM network shows a heavy-tailed degree distribution. Only a few addresses have very high in/out-degrees, while most addresses have low degrees. For instance, more than $10^5$ addresses have only one in-degree or out-degree. The low connectivity of the majority of addresses means that the network is overall sparse. However, there are only $21$ addresses that have an in-degree or out-degree of more than $10^4$, suggesting the existence of hub-like addresses in the center of the network's communication. In addition, we also observe that both the in-degree (blue line) and out-degree (red line) distributions follow a similar power-law trend on a log-log scale.

\begin{figure}[htb]
\centering
\includegraphics[width=\columnwidth]{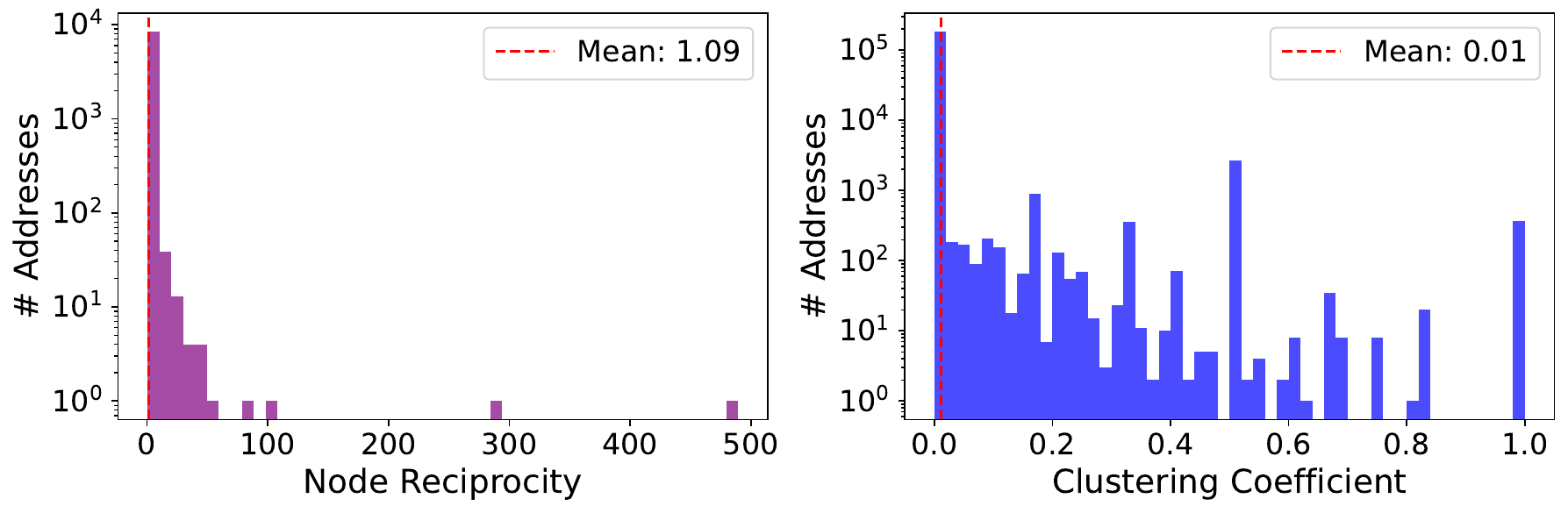}
\caption{Weighted reciprocity and clustering coefficient.}
\label{fig:Clustering_and_Reciprocity}
\end{figure}

\smallskip
\noindent\textbf{Weighted Reciprocity Distribution.} 
The left subfigure of 
Figure~\ref{fig:Clustering_and_Reciprocity} shows the distribution of the weighted reciprocal of the Ethereum \IDM network. We observe that most addresses exhibit low weighted reciprocity (with an average value of $1.09$). This indicates that IDM transactions are largely one-directional. And only $8{,}509$ ($4.50\%$) addresses engage in  \IDM communication via bidirectional transactions. This phenomenon of asymmetry indicates that users tend to leverage IDMs to broadcast information, rather than as a medium for maintaining two-way conversations.

\smallskip
\noindent\textbf{Clustering Coefficient Distribution.} 
The right subfigure of Figure~\ref{fig:Clustering_and_Reciprocity} depicts the clustering coefficient distribution of the Ethereum IDM network. We observe that the network shows a very low average clustering coefficient, i.e., $0.01$. This indicates that the network has a sparse local structure, where neighbors of an address tend not to be connected with others. Only a small number of addresses (approximately $300$-$400$) show high clustering coefficients ($\geq 0.8$), which suggests the existence of a few tightly connected local communities. In the following subsection, we will further analyze the communities of the Ethereum IDM network.

\subsection{Community Analysis}
We adopt the Louvain algorithm~\cite{blondel2008fast} to analyze the community structure of Ethereum IDM users. To analyze the meaningful interactions, we focus on the \ConnectedAddrs addresses that send messages to others, i.e., with message connections to other addresses. 

\begin{figure}[t]
\centering
\includegraphics[width=\columnwidth]{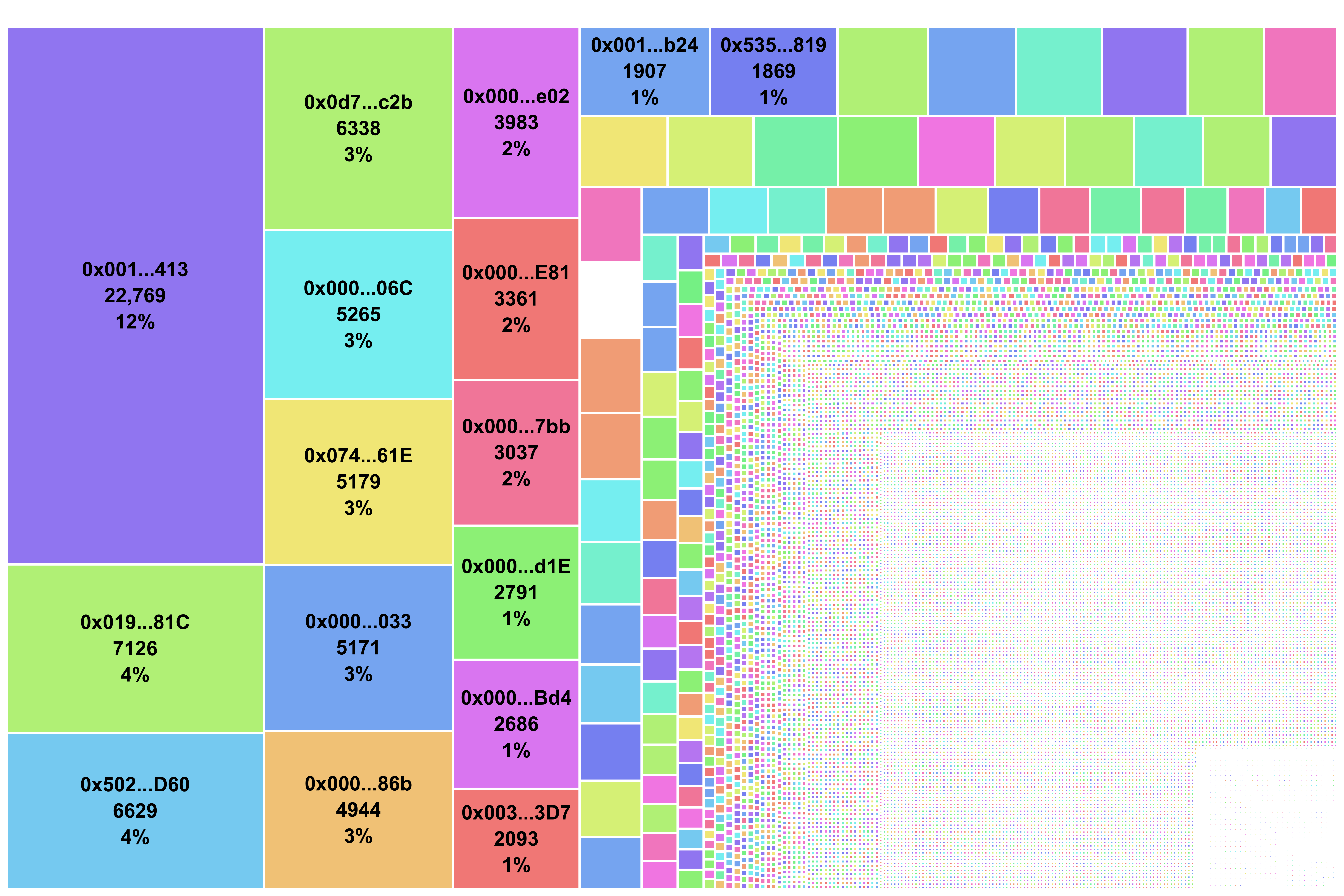}
\caption{Community distribution of Ethereum IDM users.}
\label{fig: community_treemap}
\end{figure}

\smallskip
\noindent\textbf{Community Size.}  From the \ConnectedAddrs connected addresses, we identify \numOfCommunity distinct communities, with an average size of $8.84$ addresses. As shown in Figure~\ref{fig: community_treemap}, the top $10$ largest communities account for $37.44\%$ of all addresses. Notably, the largest community contains $22{,}769$ ($12.05\%$) addresses, issuing $147{,}332$ ($34.9\%$) \IDM transactions. Interestingly, we observe that most communities ($15{,}625$, $59.99\%$) consist of only two addresses, while they only collectively account for $31{,}250$ addresses ($16.53\%$).

\begin{figure}[t]
\centering
\includegraphics[width=\columnwidth]{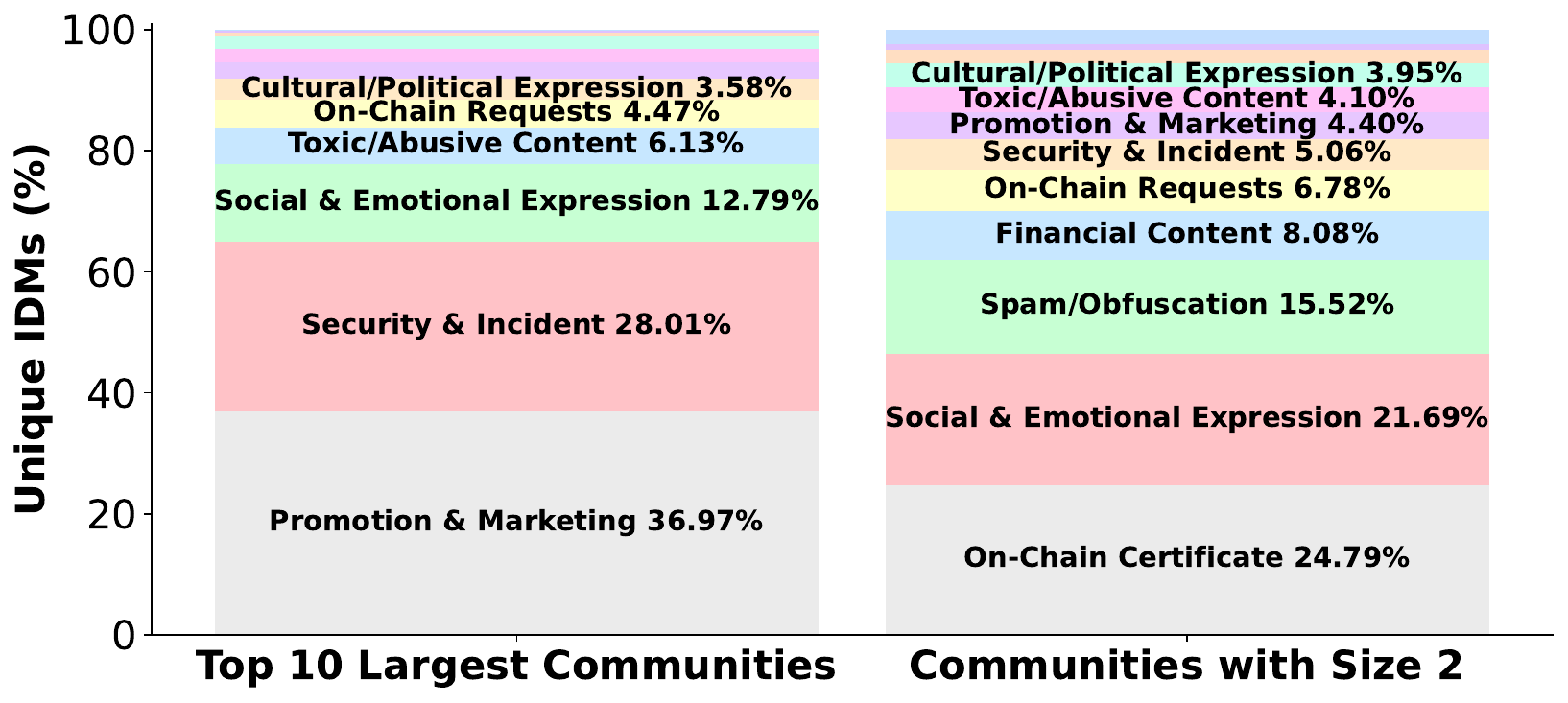}
\caption{Topic distributions in different communities.}
\label{fig: topic_distribution_comparison}
\end{figure}

\smallskip
\noindent\textbf{Community Topics.} To better understand the nature of interactions within different communities, we analyze the message topics across two distinct community types: the top $10$ largest communities and the $15{,}205$ smallest communities of exactly two addresses. Figure~\ref{fig: topic_distribution_comparison} depicts the topic distributions in these communities.

The most prevalent topics of unique IDMs issued by the top $10$ largest communities are \textit{Promotion \& Marketing} ($36.97\%$) and \textit{Security \& Incidents} ($28.01\%$).  This distribution reflects a dominant use of IDMs to disseminate promotional content or alert messages to a broad audience.
For instance, in the second-largest community, the address \hlhref{https://etherscan.io/idm?addresses=0x7a01b95c2e232d250db9e106dcf317e29a1279ab\%2C}{0x7A0...9ab} sent out at least $3{,}700$ transactions containing a message ``Check our contract scanner for more information...'', which is classified under the topic of \textit{Promotion \& Marketing}.

In contrast, the $15{,}625$ dyadic communities, which involve only two addresses, exhibit a broader range of topics. While \textit{On-Chain Certificate} represents the largest topic ($24.79\%$) in the number of unique \IDM texts, other categories emerge with significant percentages, such as \textit{Social \& Emotional Expression} ($21.69\%$), \textit{Spam/Obfuscation} ($15.52\%$), \textit{Financial Content} ($8.08\%$), etc. These topics show more diverse and personal interactions.  For instance, we observe numerous IDMs with the topic of \textit{Social \& Emotional Expression}, where users share personal thoughts, express emotions, and send greetings to their recipients. This phenomenon suggests that users in smaller communities tend to use IDMs for more personal or ideological purposes, rather than for commercially driven motives.

\vspace{-0.1in}
\begin{figure}[!]
\centering
\includegraphics[width=\columnwidth]{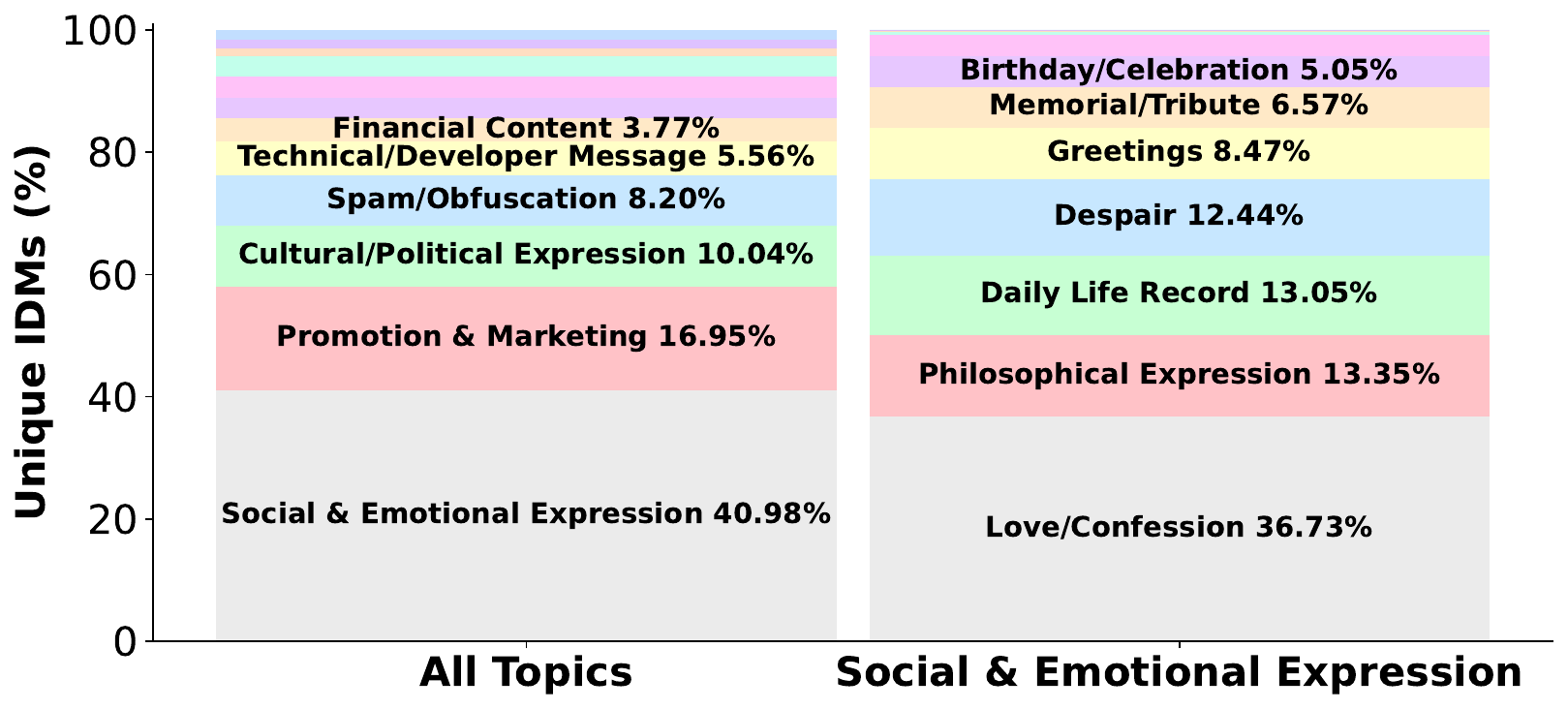}
\caption{Topic distributions for self-messaging IDMs.}
\label{fig:self_directed_topics_distribution}
\end{figure}

\subsection{Self-Messaging Behavior}
Besides interacting with others, we also identify \SelfSentAddrs addresses that send \SelfSentTxs messages to themselves. Although those addresses do not form a group or community with others, they still exhibit an interesting  ``self-messaging'' behavior.

Figure~\ref{fig:self_directed_topics_distribution} shows that self-messages span a wide range of topics. In terms of the number of unique texts over all topics, \textit{Social \& Emotional Expression} dominates at $40.98\%$. This indicates that many self-directed users tend to leverage \IDM for personal logging, signaling, or expressive purposes. We further analyze the subtopic distribution under the main topic of \textit{Social \& Emotional Expression}. We find that the top three subtopics are \textit{Love/Confession} ($36.73\%$), \textit{Philosophical Expression} ($13.35\%$), and \textit{Daily Life Record} ($13.05\%$). 

The self-messaging behavior is particularly notable among Chinese IDM senders. We highlight two illustrative examples to explore the potential motivations behind this practice. Specifically, the addresses {0x078...2b8} and {0x5cB...8fB}\footnote{Although IDMs are publicly accessible on-chain, we remove IDM links for these two examples, as they may contain user nicknames or other semi-identifiable information.} make repeated use of the IDMs to send messages to themselves. These IDMs take the form of romantic confessions and aim to publish messages on behalf of others. This behavior closely resembles the practice of ``biǎo bái qiáng'', a cultural phenomenon popular among Chinese youth.  It is conceptually similar to ``{confession page}'' on Web2 platforms like Instagram, where users share unreciprocated sentiments in a public setting. These self-directed IDMs are not intended for communication, but rather to permanently record personal messages on-chain.

\section{Security Implications}
\label{sec-security}

Interestingly, we find that many IDMs are relevant to \textit{Security \& Incident}. This phenomenon is particularly relevant in the context of emerging attack incidents on Ethereum~\cite{zhou2023sok}.

\begin{figure}[htb]
\centering
\includegraphics[width=\columnwidth]{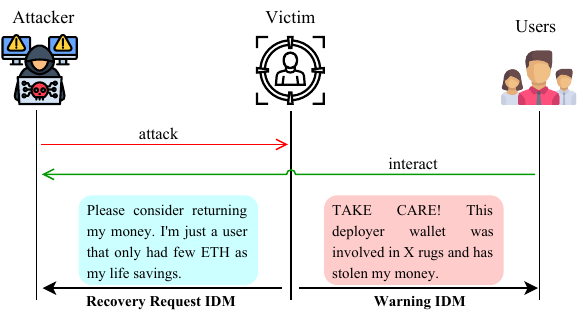}
\caption{Example of security-related IDMs.}
\label{fig:security_IDM_flow}
\end{figure}

\noindent We have identified three subtopics under \textit{Security \& Incident}:
\begin{packeditemize}
    \item \emph{Warnings ($\mathit{13{,}415}$)}~: Warnings on phishing attempts, scams, rug pulls, and other malicious activities.
    \item \emph{Attack-related ($\mathit{5{,}983}$)}~: Fund recovery requests, legal threats, or bounty offers following security breaches.
    \item \emph{Public Apology ($\mathit{137}$)}~: Apologies issued by protocols or teams in response to incidents or failures.
\end{packeditemize}

Figure~\ref{fig:security_IDM_flow} provides examples of attack-related IDMs. A victim in an accident may request the attacker to return money, and simultaneously warn other users of the associated risks.

\subsection{IDMs with Security Relevance}
\label{sec: idm_security}

\begin{figure}[htb]
\centering
\includegraphics[width=\columnwidth]{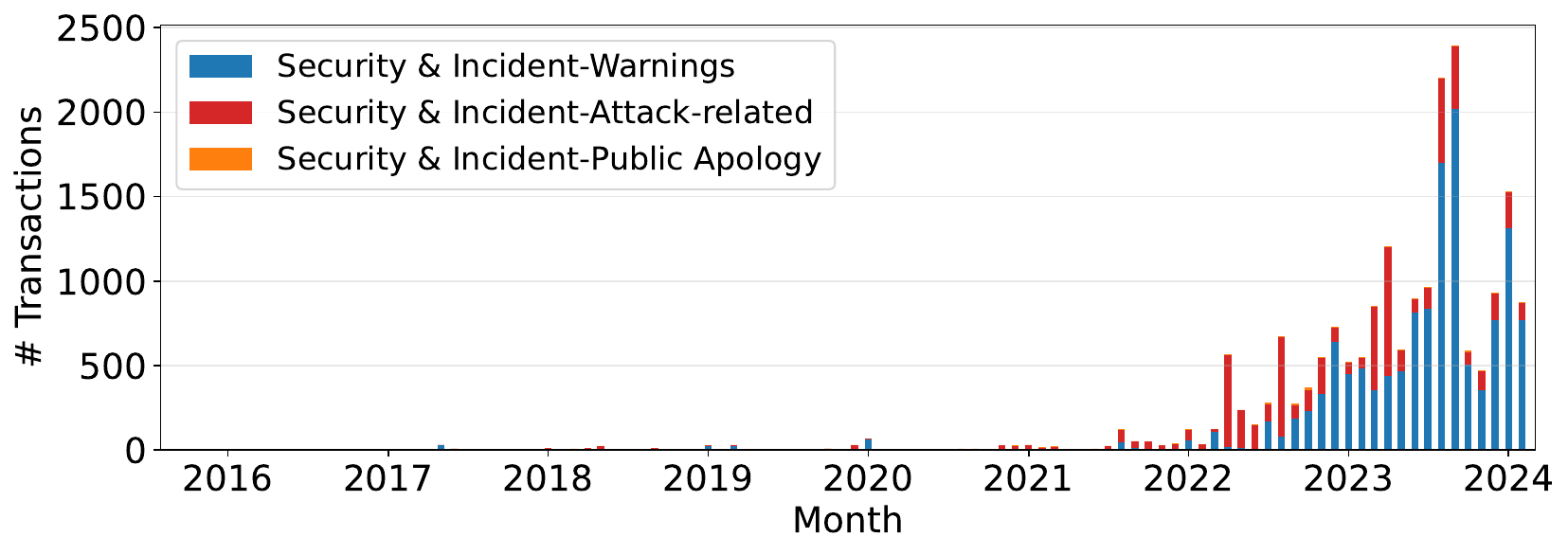}
\caption{Temporal trends of IDMs with security relevance.}
\label{fig:security_topics_evolution_monthly}
\end{figure}

As shown in Figure~\ref{fig:security_topics_evolution_monthly}, the volume of IDMs with security relevance has grown significantly since 2022. This is probably due to the explosive growth of DeFi. Before 2022, the volume for all three subtopics was negligible. This is likely due to the relatively small scale of the DeFi ecosystem and low exposure to on-chain attacks.

The spike volume of the \textit{Warnings} subtopic appears in 2023. As shown in Table~\ref{tab:topic_taxonomy}, these warning IDMs typically relate to phishing scams, rug pulls, and other malicious activities. These activities have become frequent as DeFi platforms rapidly onboard users and assets without centralized regulation. In response, affected users or victims could turn to on-chain \IDMs to share alerts and raise awareness among other users. For instance, as shown in Figure~\ref{fig:security_IDM_flow}, users sent messages such as ``TAKE CARE! This deployer wallet was involved in X rugs'' to their peers. This indicates that IDMs can serve as a peer-to-peer warning function to achieve a collective safeguard mechanism within the Ethereum community.

\begin{figure}[htb]
\centering
\begin{minipage}{.22\textwidth}
  \centering
  \includegraphics[width =\textwidth]
  {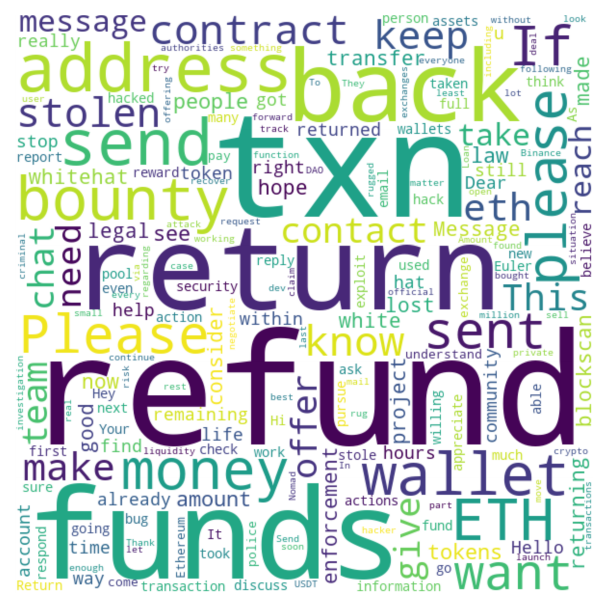}
\end{minipage}%
\hfill
\begin{minipage}{.22\textwidth}
  \centering
  \includegraphics[width =\textwidth]
  {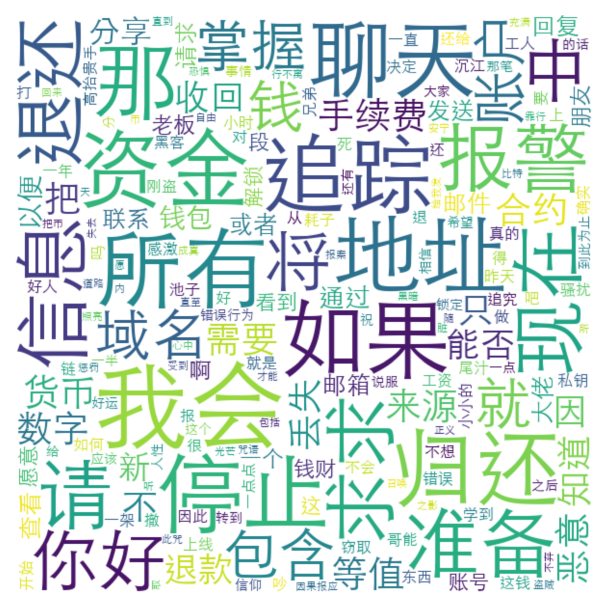}
\end{minipage}
\caption{English (left) and Chinese (right) IDM word clouds for the \textit{Security \& Incident} main topic.}
\label{fig:security_ch_cn_popular_word}
\end{figure}

We also observe a similar trend in attack-related IDMs, which typically involve fund recovery requests, legal threats, or bounty offers after security breaches. As shown in Figure~\ref{fig:security_ch_cn_popular_word}, common keywords in these IDMs, such as ``return'', ``refund'', and ``fund'', frequently appear in both English and Chinese. Using LLMs, we identify the following strategies that users adopt to communicate with the attackers for fund recovery requests: 

\begin{packeditemize}
    \item \emph{Plead}: The sender appeals to the attacker by highlighting personal hardship or financial distress.
    \item \emph{Threaten}: The sender warns that the attacker's identity has been exposed and threatens to involve regulators.
    \item \emph{Reward}: The sender offers incentives, such as bounties or a share of the stolen funds, in exchange for their return.
    \item \emph{Negotiate}: The sender proposes a compromise, such as partial repayment or a mutually acceptable arrangement.
\end{packeditemize}

\begin{table}[htb]
\centering
\caption{Communication strategies and effectiveness.}
\label{tab:security_approaches}
\vspace{-0.08in}
\renewcommand\arraystretch{1}
\resizebox{0.9\linewidth}{!}{
\begin{tabular}{r|ccc}
\toprule
\multicolumn{1}{c}{\textbf{Strategies}} & \# Unique IDM & \# Reply & Reply Rate (\%) \\
\midrule
Plead \includegraphics[height=0.95em]{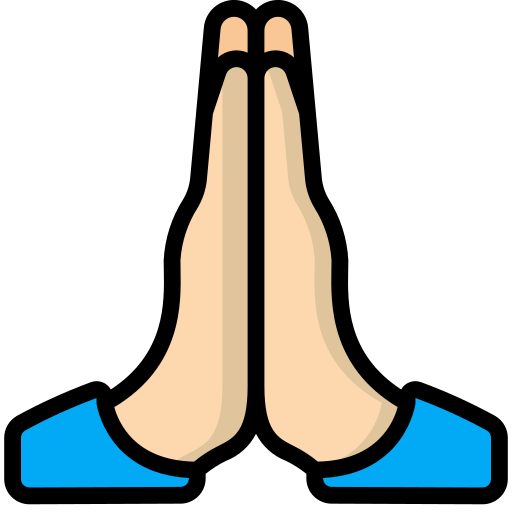} & 287 & 21 & 7.3\% \\
Threaten \includegraphics[height=0.93em]{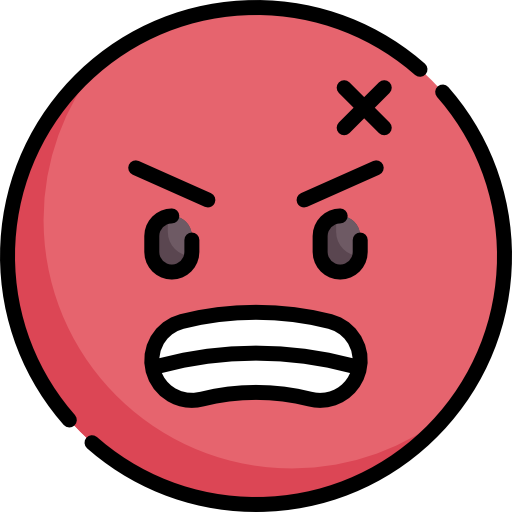}  & 599 & 32 & 5.3\% \\
Reward \includegraphics[height=1em]{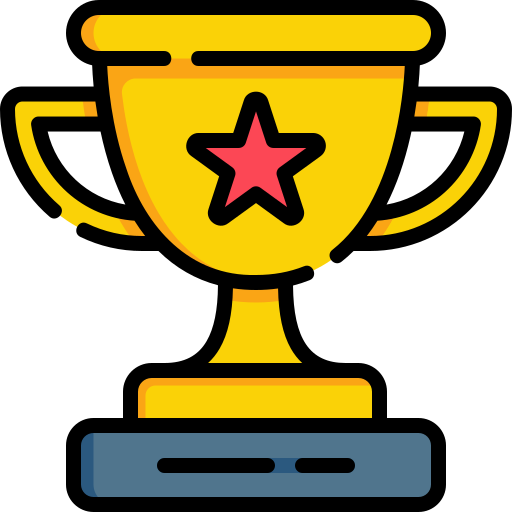} & 245 & 39 & \cellcolor{red!15}15.9\% \\
Negotiate \includegraphics[height=0.93em]{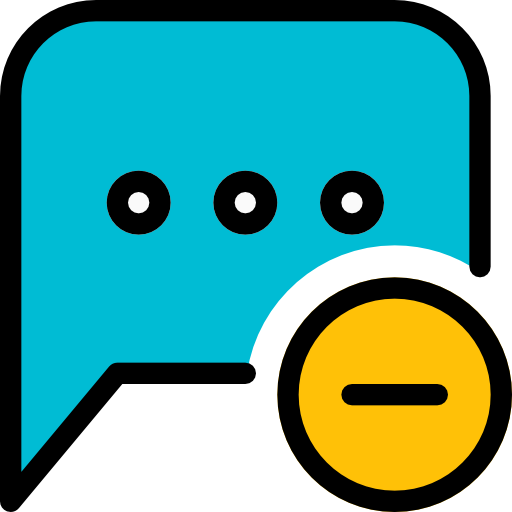} & 485 & 95 & \cellcolor{red!15}19.6\% \\
\bottomrule
\end{tabular}
}
\end{table}

To evaluate the effectiveness of different fund recovery strategies, we analyze whether attackers return the stolen funds to users. As shown in Table~\ref{tab:security_approaches}, recovery requests that involve rewards or negotiation are notably more successful, achieving reply rates of $15.9\%$ and $19.6\%$, respectively. In contrast, strategies based on pleading or threatening are less effective, and the threat approach has the lowest reply rate at merely $5.3\%$.  The result suggests that appeals involving legal threats or regulatory involvement are largely unconvincing. This is likely due to the absence of regulatory frameworks and enforcement mechanisms in the Ethereum ecosystem.

\subsection{Overlap with On-Chain Mixer Users}
To further understand security-related IDM user behaviors, we investigate their intersection with users of Tornado Cash (TC)~\cite{wang2023zero}. TC is a widely used on-chain mixer designed to enhance users' privacy by breaking the linkability between addresses. Engaging with TC typically suggests an awareness of preserving privacy.

\smallskip
We first crawl the depositor addresses in TC ETH pools from block $9{,}117{,}019$ (December 16, 2019) to block $21{,}998{,}172$ (March 07, 2025), identifying a total of $48{,}279$ unique depositors.
Table~\ref{tab:tc_overlap} presents the overlap between IDM participants (both senders and receivers) and TC depositors, categorized by IDM topics. Notably, we observe that users involved in the \textit{Security \& Incident: Attack-related} topic exhibit the highest overlap with TC usage, i.e., $5.2\%$ of senders and $5.9\%$ of receivers. This indicates a correlation between security-related communication and privacy-seeking behavior.  We suspect that users who communicate about security incidents are more likely to have sensitive operational motives and, correspondingly, are more inclined to utilize privacy-enhancing solutions.

In contrast, IDM users engaged in more outward-facing topics, such as \textit{Promotion \& Marketing} or \textit{Education}, show minimal overlap with TC users. For example, users associated with the \textit{Promotion \& Marketing} topic show a Tornado Cash usage rate of less than $1\%$. This further suggests that TC is more commonly used by IDM users engaged in communication related to risk-associated topics.

\begin{table}[htbp]
\centering
\caption{Overlap between IDM users and TC depositors.}
\label{tab:tc_overlap}
\vspace{-0.1in}
\renewcommand\arraystretch{1}
\resizebox{\linewidth}{!}{
\begin{tabular}{l|rrrrr}
\toprule
\multicolumn{1}{c}{\textbf{Topics}} & \makecell{\# Unique\\ Senders} & \makecell{\# Senders\\ TC Overlap(\%)} & \makecell{\# Unique\\ Receivers} & \makecell{\# Receivers\\ TC Overlap(\%)} \\
\toprule
\rowcolor{pink!75}
Security \& Incident-Attack-related & 1,488 & 78 (5.2\%) & 4,505 & 264 (5.9\%) \\

Security \& Incident-Warnings & 998 & 22 (2.2\%) & 10,439 & 88 (0.8\%) \\

Security \& Incident-Public Apology & 106 & 3 (2.8\%) & 119 & 6 (5.0\%) \\

\midrule
Social \& Emotional Expression & 6,239 & 152 (2.4\%) & 9,139 & 140 (1.5\%) \\
Promotion\&Marketing & 10,989 & 40 (0.4\%) & 34,368 & 156 (0.5\%) \\
Spam/Obfuscation & 35,466 & 54 (0.2\%) & 48,430 & 41 (0.1\%) \\
On-Chain Requests & 3,325 & 89 (2.7\%) & 40,181 & 291 (0.7\%) \\
On-Chain Certificate & 1 & 0 (0.0\%) & 1 & 0 (0.0\%) \\
Cultural/Political Expression & 1,509 & 20 (1.3\%) & 1,408 & 49 (3.5\%) \\
Financial Content & 1,627 & 39 (2.4\%) & 8,330 & 105 (1.3\%) \\
Toxic/Abusive Content & 1,308 & 32 (2.4\%) & 2,353 & 52 (2.2\%) \\
Technical/Developer Message & 1,554 & 30 (1.9\%) & 1,954 & 35 (1.8\%) \\
Education & 298 & 3 (1.0\%) & 146 & 2 (1.4\%) \\
Charity/Fundraising & 165 & 3 (1.8\%) & 234 & 8 (3.4\%) \\

\bottomrule
\end{tabular}
}
\end{table}

\section{Moderation and Regulation Implications}
\label{sec-regulation}

This section examines toxic \IDMs to show how IDMs can carry harmful content and why content moderation and regulatory attention are necessary within decentralized environments.

\subsection{Analysis of Toxic/Abusive IDMs} \label{sec: toxic_semantic}

\begin{figure}[htb]
\centering
\includegraphics[width=\columnwidth]{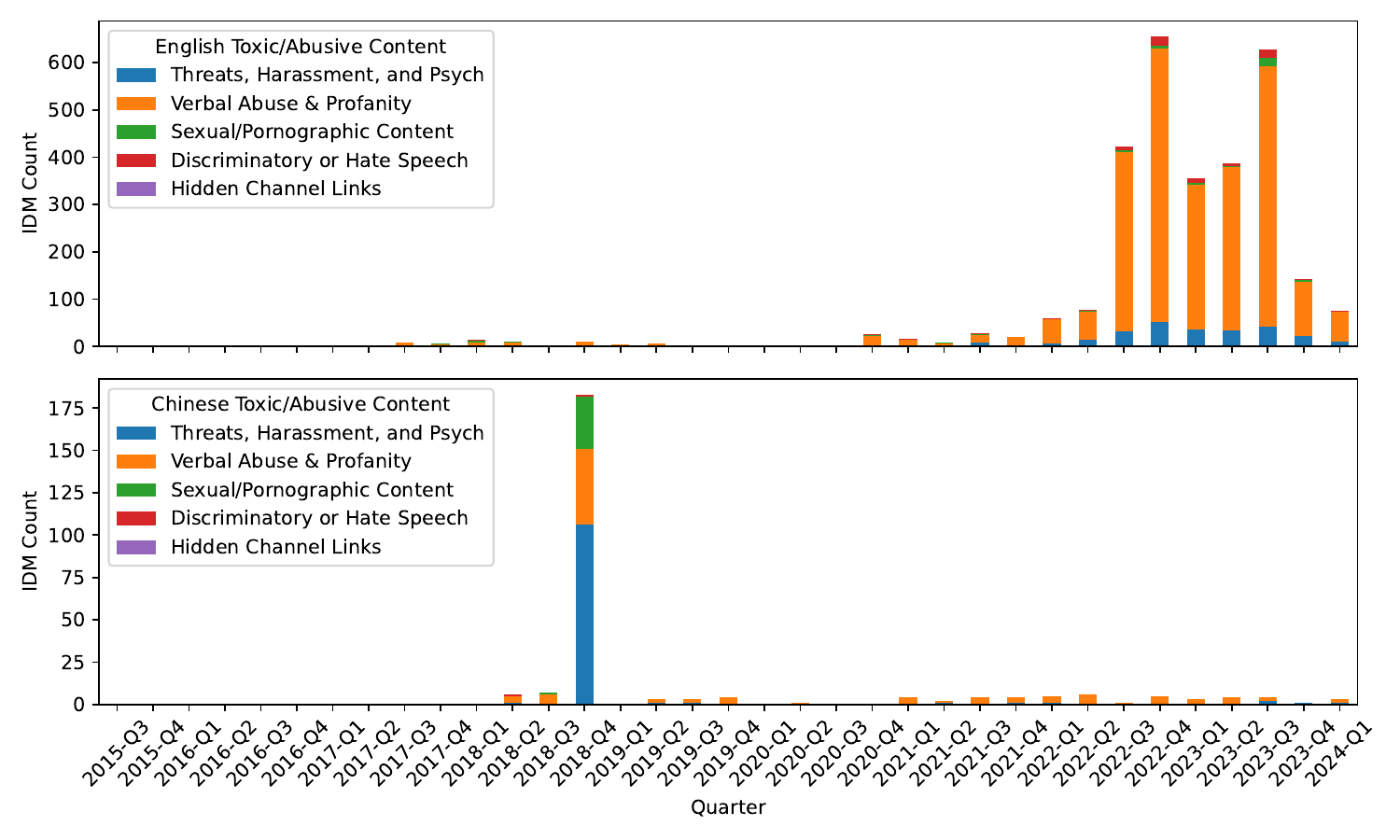}
\vspace{-0.15in}
\caption{Temporal trends of toxic IDMs by subtopics in English (top) and Chinese (bottom).}
\label{fig:toxic_cn_en_topic_over_time}
\vspace{-0.1in}
\end{figure}

Figure~\ref{fig:toxic_cn_en_topic_over_time} compares the temporal dynamics of toxic or abusive content in Ethereum \IDMs between English and Chinese messages. English \IDMs show a clear rise in toxic content starting from late $2020$. This trend accelerates throughout $2022$ and $2023$. Most of the content falls under the topic of \textit{Verbal Abuse \& Profanity}. Smaller portions relate to threats or psychological harassment. A few messages contain hate speech or sexual references.

In contrast, toxic content in Chinese \IDMs is minimal. It appears only occasionally over time, with one sharp but isolated peak in $2018$. Beyond that, the frequency remains low and stable. This contrast suggests that toxic expression on-chain is more common in English-language communication. Chinese-language messages show limited use of IDMs for antagonistic or abusive discourse.

\begin{table}[htbp]
\centering
\caption{Emotional profiles of toxic IDM subtopics.}
\label{tab: toxic_emotion}
\vspace{-0.08in}
\resizebox{\columnwidth}{!}{%
\begin{tabular}{l|ccccc}
\toprule
\multicolumn{1}{c}{\textbf{Emotion}} & \textbf{Hate Speech} & \textbf{Hidden Links} & \textbf{Sexual} & \textbf{Threats} & \textbf{Verbal Abuse} \\
\midrule
Negative-Anger & \cellcolor{red!0}2 & \cellcolor{red!0}1 & \cellcolor{red!0}0 & \cellcolor{red!0}18 & \cellcolor{red!15}312 \\
Negative-Disgust & \cellcolor{red!0}11 & \cellcolor{red!0}0 & \cellcolor{red!1}21 & \cellcolor{red!1}34 & \cellcolor{red!10}219 \\
Negative-Fear & \cellcolor{red!0}0 & \cellcolor{red!0}0 & \cellcolor{red!0}6 & \cellcolor{red!3}74 & \cellcolor{red!0}20 \\
Negative-Guilty & \cellcolor{red!0}0 & \cellcolor{red!0}0 & \cellcolor{red!0}0 & \cellcolor{red!0}0 & \cellcolor{red!0}1 \\
Negative-Hostility & \cellcolor{red!3}63 & \cellcolor{red!0}0 & \cellcolor{red!0}11 & \cellcolor{red!10}221 & \cellcolor{red!50}2,044 \\
Negative-Sadness & \cellcolor{red!0}0 & \cellcolor{red!0}0 & \cellcolor{red!0}8 & \cellcolor{red!1}31 & \cellcolor{red!0}16 \\
\hline
Neutral-Confusion & \cellcolor{red!0}1 & \cellcolor{red!0}0 & \cellcolor{red!0}1 & \cellcolor{red!0}0 & \cellcolor{red!0}12 \\
Neutral-Curiosity & \cellcolor{red!0}0 & \cellcolor{red!0}1 & \cellcolor{red!0}5 & \cellcolor{red!0}2 & \cellcolor{red!0}12 \\
Neutral-Politeness & \cellcolor{red!0}0 & \cellcolor{red!5}116 & \cellcolor{red!0}3 & \cellcolor{red!0}0 & \cellcolor{red!0}10 \\
Neutral-Surprise & \cellcolor{red!0}0 & \cellcolor{red!0}0 & \cellcolor{red!0}3 & \cellcolor{red!0}0 & \cellcolor{red!0}0 \\
\hline
Positive-Gratitude & \cellcolor{red!0}0 & \cellcolor{red!0}0 & \cellcolor{red!0}0 & \cellcolor{red!0}0 & \cellcolor{red!0}4 \\
Positive-Hope & \cellcolor{red!0}0 & \cellcolor{red!0}0 & \cellcolor{red!0}0 & \cellcolor{red!0}1 & \cellcolor{red!0}0 \\
Positive-Joy & \cellcolor{red!0}2 & \cellcolor{red!0}0 & \cellcolor{red!0}5 & \cellcolor{red!0}1 & \cellcolor{red!0}6 \\
Positive-Trust & \cellcolor{red!0}0 & \cellcolor{red!0}0 & \cellcolor{red!0}0 & \cellcolor{red!0}2 & \cellcolor{red!0}5 \\
Positive-Warmth & \cellcolor{red!0}0 & \cellcolor{red!0}0 & \cellcolor{red!0}2 & \cellcolor{red!0}0 & \cellcolor{red!0}3 \\
\bottomrule
\end{tabular}%
}
\begin{tablenotes}[flushleft]
      \footnotesize
      \item[] \quad \textit{Color intensity indicates count magnitude.}
\end{tablenotes}
\end{table}

Table~\ref{tab: toxic_emotion} breaks down toxic \IDM subtopics by emotion category. Most toxic content is associated with negative emotions, especially \textit{Hostility} and \textit{Anger}. \textit{Hostility} appears most frequently, dominating both \textit{Verbal Abuse \& Profanity} and \textit{Threats, Harassment, and Psych}. \textit{Disgust} also shows a strong link to \textit{Verbal Abuse \& Profanity} and \textit{Sexual/Pornographic Content}. Surprisingly, a small number of polite messages are tied to \textit{Hidden Channel Links}, suggesting deceptive intent. These patterns highlight how certain emotions are closely tied to abusive or manipulative behaviors on-chain.

\subsection{Case Study of Toxic/Abusive IDMs}
\label{sec: case_study}
We present some case studies to show how the misuse of Ethereum \IDMs can facilitate the propagation of toxic information\footnote{Claim: The toxic IDMs cited in this paper are included solely for the purpose of analysis. We do not endorse or condone their content in any form. All excerpts are presented to support research on IDM misuse and the need for governance and regulation. For this discussion, IDM links are removed to comply with the Anti-Harassment Policy.}.

\smallskip
\noindent\textbf{Verbal Abuse \& Profanity.} 
Verbal abuse and profanity constitute the most prevalent form of toxic IDMs. They are characterized by aggressive tones and vulgar expressions. One example appears in transaction~{0xc36...7f6}, where the sender directs a violently profane message. It includes crude language and an explicit invocation of violence, reflecting a clear intent to insult and intimidate. This case demonstrates how \IDMs can be misappropriated for targeted abuse.

\smallskip
\noindent\textbf{Threats, Harassment, and Psych.} 
Threats and harassment are among the most severe forms of abuse observed in IDMs. In certain cases, messages go beyond profanity and contain explicit death threats, often directed at specific individuals or groups. One striking example appears in~{0x9a94...de4}, where the sender issues repeated threats against an entire family, using hostile and violent language. This example reflects the alarming potential for on-chain messages to be weaponized for psychological harm. The public visibility of blockchain storage amplifies the impact of such messages.

\smallskip
\noindent\textbf{Discriminatory/Hate Speech.} 
Discriminatory or hate speech represents one of the most extreme abuses of the \IDM channel. In certain cases, senders have injected explicit racial and ethnic slurs directly into the input data field, leveraging the immutability of the blockchain to preserve hostile speech. A disturbing example can be found in transaction~{0x6f31...4b2}, where the message consists entirely of hate-filled language targeting multiple identity groups. This case illustrates how on-chain messaging can be misused to propagate violent ideologies.

\smallskip
\noindent\textbf{Sexual/Pornographic Content.}
Some IDMs contain unsolicited sexual language or explicit propositions, reflecting the misuse of input data as a vector for inappropriate or predatory communication. One such case can be seen in transaction~{0xbb68...07c}, where the sender issues a crude sexual request in broken English. The language is explicitly directed toward the recipient, indicating both an intent to provoke and a disregard for any communicative norms.

This type of content mirrors patterns found in unsolicited sexual messaging across web2 platforms, but the immutable nature of blockchain messaging means such content cannot be removed. 

\smallskip
\noindent\textbf{Hidden Channel Links.}
Some IDMs embed external links that lead to off-chain communication spaces. These messages typically use short, informal phrases to draw attention, followed by a link to a messaging group, redirect service, or promotional site. For instance, one message promotes a ``hidden mint'' event through a Telegram message link ({0xe310...e28}),  which appears to be part of a token or NFT marketing strategy.  Another example ({0xd597...381}) uses sexual language and directs users to a suspicious-looking URL, likely for adult content or scam purposes. These messages are often short and distributed across many recipients, resembling patterns of phishing campaigns. While they do not always contain direct abuse, their intent is often exploitative.

\subsection{Implications for Moderation and Regulation}

The case studies presented in \S\ref{sec: case_study} demonstrate that blockchain messaging can be misused for severe forms of verbal abuse, targeted threats, hate speech, sexual solicitation, and spam-related link propagation. They reflect how a decentralized system can also serve as a channel for hostile, exploitative, and manipulative communication.

\begin{table}[tbh]
    \centering
    \caption{Content moderation in Web2 and Web3 platforms.}
    \label{tab: content_moderation}
    \vspace{-0.08in}
    \renewcommand\arraystretch{1.2}
    \resizebox{\columnwidth}{!}{
    \begin{tabular}{c|cccccc}
    \toprule
    \textbf{Platform} & \makecell[c]{Moderated\\Content} & \makecell[c]{AI-Based\\Moderation}	 & \makecell[c]{Human\\Review} & 	\makecell[c]{User\\Reporting}	& \makecell[c]{Enforcement\\Actions}\\
     \midrule
    \makecell[c]{Facebook\\(Meta)} & \makecell[l]{
    - Hate speech \& harassment\\
    - Violent or graphic content\\
    - Nudity \& sexual content\\
    - Misinformation and spam\\
    - Terrorism \& illegal activity} & \cmark & \cmark & \cmark &  \makecell[l]{
    - Content removal\\
    - Content down-ranking \\
    - Fact-check labels\\
    - Temporary bans\\
    - Account suspensions
    } \\
    
    \cmidrule{4-6}
    
    TikTok & \makecell[l]{
    - Hate speech \& harassment\\
    - Violent or graphic content\\
    - Nudity \& sexual content\\
    - Self-harm \& dangerous acts\\
    - Misinformation \& crime
    }& \cmark & \cmark & \cmark & \makecell[l]{
    - Content removal\\
    - Content down-ranking \\
    - Account warnings \\
    - Temporary bans\\
    - Account suspensions
    } \\
    
   \midrule
    
    \makecell[c]{Memo.Cash\\(Web3)} & -  & \xmark & 
    \xmark & \xmark & - \\

    \cmidrule{4-6}
    
    \makecell[c]{Ethereum \\ IDM} & -  & \xmark & 
    \xmark & \xmark & - \\
    \bottomrule
    \end{tabular}
    }
\end{table}

In Web2 environments, such content would typically be removed under the platform's content moderation system~\cite{gorwa2020algorithmic}. For example, mainstream social media such as Facebook\footnote{\href{https://transparency.meta.com/en-gb/policies/community-standards/}{Community Standards}, Meta Transparency Center.} and TikTok\footnote{\href{https://www.tiktok.com/community-guidelines/en}{Community Guidelines}, TikTok.} explicitly prohibit toxic information such as threats, hate speech, and sexual content, using a combination of AI-based and human moderation (see Table~\ref{tab: content_moderation}). These systems are supported by centralized authority and reactive enforcement, allowing for timely interventions.

In contrast, blockchain ecosystems offer no built-in moderation mechanisms. Once deployed, input data is immutable, and there are no platform-level controls to report or remove harmful content. Messages are not bound by shared community guidelines, and senders remain pseudonymous and unaccountable. Cases like \href{https://memo.cash/}{Memo.cash} show the risks of decentralized social media~\cite{zuo2024understanding}. The platform, built on Bitcoin Cash, has hosted hate speech and offensive posts, with no effective way to moderate them\footnote{Memo.cash implements a user-level \textit{mute} mechanism, allowing individuals to filter unwanted content. However, this is not a form of content moderation in the traditional sense, as the content remains publicly visible and immutable on the blockchain.}. This challenge extends to Ethereum IDMs. As our study shows, Ethereum has already seen the use of IDMs to propagate toxic and abusive content. Yet, there is no effective moderation layer. This raises concerns about the long-term societal impact of unmoderated communication in decentralized systems. The need for regulatory efforts and governance mechanisms is becoming increasingly urgent.

Furthermore, Web2 platforms are often operated in regulated settings. For example, the EU’s Digital Services Act and China’s Cybersecurity Law require platforms to take responsibility for harmful or illegal posts. These rules mandate content removal, reporting, and compliance mechanisms. In contrast, decentralized systems operate without such regulatory frameworks. There are no legal duties for moderation, and no clear paths for accountability. This legal gap presents serious challenges for harm prevention.

\section{Ethical Considerations}
\label{sec-ethical}

To mitigate ethical concerns, we present the following claims.

\vspace{0.6mm}
\noindent\textbf{Public Data Sources and Respect for Law.} 
Our work presents the first large-scale empirical study of IDMs on the Ethereum mainnet. All data originates from a public, permissionless ledger, and our analysis involves only the interpretation of this openly available information. Our data practices comply with GDPR Recital 26 (data manifestly made public), as well as relevant data protection laws in the countries where the authors are based. No smart-contract vulnerabilities were exploited, and no transaction was emitted by the research team. All IDMs involved in this paper are included solely for the purpose of academic research. We do not endorse or condone their content in any form.

\vspace{0.6mm}
\noindent\textbf{Privacy Considerations.}  
All data analyzed in this study consists solely of on-chain transactions, collected from a self-hosted Ethereum archive node. No off-chain or auxiliary datasets were used at any point.  Although these records are already public and pseudonymous, we applied an additional \emph{privacy-hardening pipeline}:

\begin{packeditemize}
    
  \item \textbf{Address Truncation.}  
        Each hexadecimal address, except those related to public incidents and events, is reduced to its first five and last three characters (e.g., {0x123...abc}) so that linkability with Web traces or APIs is weakened. 

  \item \textbf{Transaction Link Suppression for Toxic IDMs.}  
        Although we include toxic or abusive IDMs in our analysis and discuss their implications for regulation and moderation in \S\ref{sec-regulation}, we remove links to transaction hashes whose IDMs \emph{(i)} contain personally identifiable or semi-identifiable information, \emph{(ii)} include or reference sexually explicit content, or \emph{(iii)} involve legally prohibited expressions of extremism, hate speech, or threats.
       
\end{packeditemize}

\vspace{0.6mm}
\noindent\textbf{Access Upon Request.}  
Researchers with a legitimate and non-commercial need (e.g., auditing the redaction procedure) may request the \textit{full} access to our crawled Ethereum \IDMs under a no-redistribution data-use agreement. 

\vspace{0.6mm}
\noindent By adhering to these principles, we aim to maximize scientific value while respecting the rights, safety, and privacy of community users.

\section{Related Work}
\label{sec-rw}

\noindent\textbf{IDM with functional usage.}  
The literature to date has examined the \inputdata field almost exclusively through a \emph{functional} lens.  Two main application families have emerged.

\begin{packeditemize}
    \item  \textit{Covert channels.}  
    A series of works~\cite{zhang2024covert,zhang2023covert} show that ordinary transactions can be used to \textit{sneak} arbitrary data onto a blockchain by piggy-backing extra bytes inside fields meant for contract parameters or script outputs, creating an uncensorable message board. On Ethereum the bytes are hidden in the \textit{input data} field~\cite{liu2020whispers,chen2022blockchain} after external encryption processes; on Bitcoin they sit in the 80-byte \texttt{OP\_RETURN} output \cite{zhu2023novel}.
    
    \item \textit{On-chain inscription.}  
    With the rise of \textit{inscription}~\cite{li2024bitcoin}, users began packing Base64-encoded JPEG/GIF fragments into IDMs to mint so-called \emph{vanilla NFTs}, i.e.\ digital artifacts that live fully on L1 without an ERC-721 contract~\cite{messias2024writing}. 
\end{packeditemize}

However, the vast body of \emph{non-functional} messages that users voluntarily embed are unexplored.  We fill the gap by providing the first analysis of IDMs as a decentralized communication channel, extending the purely utilitarian perspective of prior studies.

\smallskip
\noindent\textbf{LLM-assisted Blockchain Analyses.}  
Large-language models have recently become a versatile lens on on-chain data.  The first wave of studies focused on \emph{smart-contract vulnerability detection} by converting Solidity byte- or source-code into natural-language–like tokens that GPT-series and code-centric LLMs (e.g., CodeBERT, CodeT5) can reason over, outperforming classic symbolic analyzers on re-entrancy, unchecked \texttt{call} and access-control flaws~\cite{chen2023chatgpt,wang2024smartinv,sun2025adversarial,ince2024detect,sun2024gptscan,ma2024combining,ge2025adapting,xiao2025logic,li2025scalm}.

Beyond code auditing, researchers applied foundation models to higher-level blockchain artifacts, including transaction-graph understanding~\cite{lei2025large}, entity labeling~\cite{avice2025linking}, regulation mappings~\cite{luo2025decoding},  code generation~\cite{liu2024propertygpt}, gas analysis~\cite{jiang2024unearthing}, and anomaly detection~\cite{avice2025linking}.

\smallskip
\noindent\textbf{On-chain Empirical Study.} Blockchains offer a uniquely rich public data source, far more open than the proprietary logs held by traditional financial platforms. A typical empirical workflow \emph{(i)} harvests raw blocks and mempool traces, \emph{(ii)} reconstructs higher-level objects (e.g., token balances), and \emph{(iii)} applies graph, econometric, or NLP techniques to extract insight. Using this recipe, prior work has mapped decentralized exchanges~\cite{han2021trust}, arbitrage dynamics~\cite{wu2025hunting,daian2020flash}, profiled NFT markets \cite{wang2023brc,messias2024writing}, crypto scams~\cite{yang2024stole,bolz2024machine}, and even ``black-swan'' events~\cite{wang2024cryptocurrency}.

Those studies treat the transaction \textit{input data} field as an opaque payload, or even prune it for space (e.g., EIP-4444~\cite{eip4444}). However, we place that field at the centre stage and conduct the first IDM analysis at scale. We uncover a user-facing communication layer that has been largely invisible to prior empirical research.

\section{Limitations and Future Work}

\noindent \textbf{Limitations.} 
Due to the unique features of IDMs and their analytical challenges (\S\ref{sec-model}), we leverage LLMs to support several tasks, such as language detection, topic and sentiment classification. While LLM outputs are sensitive to prompt design and model behavior~\cite{zhou2024llm,zhong2024can}, our prompts are carefully designed and constructed. In addition, we involve human review to ensure classification consistency. We do not treat the LLM as ground truth, but rather as an assisting tool for human-guided analysis.

\smallskip
\noindent \textbf{Future Work.}
Future work may extend our LLM-assisted analysis by developing more robust and scalable pipelines. This may include model fine-tuning and cross-model validation. Building a labeled dataset from \IDMs may also enable training domain-adapted classifiers for more systematic analysis. 
Additionally, our findings reveal that Ethereum lacks a built-in content moderation infrastructure for toxic \IDMs. Future work could explore decentralized moderation designs to mitigate \IDM misuse and achieve harm prevention.

\section{Conclusion}

By adopting a ``transaction-as-communication'' perspective, this paper presents the first large-scale, systematic analysis of Ethereum Input Data Messages (IDMs).  Our analysis reveals cross-cultural divergences, emotional structures, functional intentions, network patterns, security implications, and governance challenges embedded in on-chain messages. We hope these findings help establish a new paradigm for understanding blockchain not only as a financial infrastructure, but as a socio-technical medium for decentralized communication. We also envision that this work will serve as a foundation for interdisciplinary research to explore and improve the social functionality of decentralized systems.

\newpage

\bibliographystyle{unsrt}
\bibliography{ref}

\begin{thebibliography}{10}

\bibitem{bitcoin}
Satoshi Nakamoto.
\newblock Bitcoin: A peer-to-peer electronic cash system.
\newblock {\em Available at: \url{https://bitcoin.org/bitcoin.pdf}}, 2008.

\bibitem{wood2014ethereum}
Gavin Wood.
\newblock Ethereum: A secure decentralised generalised transaction ledger.
\newblock {\em Ethereum project yellow paper}, 151:1--32, 2014.

\bibitem{antonopoulos2018mastering}
Andreas~M Antonopoulos and Gavin Wood.
\newblock {\em Mastering ethereum: building smart contracts and dapps}.
\newblock O'reilly Media, 2018.

\bibitem{idm2023}
Teck~Yuan Lee.
\newblock Etherscan information center: Understanding transaction input data.
\newblock {\em Retrieved by May 2025,
  \url{https://info.etherscan.com/understanding-transaction-input-data/}},
  2023.

\bibitem{bybit2025}
TRM Labs.
\newblock The {B}ybit hack: Following {N}orth {K}orea’s largest exploit.
\newblock {\em Retrieved by May 2025,
  \url{https://www.trmlabs.com/resources/blog/the-bybit-hack-following-north-koreas-largest-exploit}},
  2025.

\bibitem{jauhiainen2019automatic}
Tommi Jauhiainen, Marco Lui, Marcos Zampieri, Timothy Baldwin, and Krister
  Lind{\'e}n.
\newblock Automatic language identification in texts: A survey.
\newblock {\em Journal of Artificial Intelligence Research}, 65:675--782, 2019.

\bibitem{werner2022sok}
Sam Werner, Daniel Perez, Lewis Gudgeon, Ariah Klages-Mundt, Dominik Harz, and
  William Knottenbelt.
\newblock {SoK}: Decentralized finance {(DeFi)}.
\newblock In {\em ACM Conference on Advances in Financial Technologies (AFT)},
  pages 30--46, 2022.

\bibitem{jiang2023decentralized}
Erya Jiang, Bo~Qin, Qin Wang, Zhipeng Wang, Qianhong Wu, Jian Weng, Xinyu Li,
  Chenyang Wang, Yuhang Ding, and Yanran Zhang.
\newblock Decentralized finance ({DeFi}): A survey.
\newblock {\em arXiv preprint arXiv:2308.05282}, 2023.

\bibitem{rodino2018me}
Michelle Rodino-Colocino.
\newblock Me too,\# metoo: Countering cruelty with empathy.
\newblock {\em Communication and critical/cultural studies}, 15(1):96--100,
  2018.

\bibitem{istabul2019}
Ethereum.
\newblock Ethereum.org: The history of ethereum.
\newblock {\em Retrieved by May 2025, \url{https://ethereum.org/en/history/}},
  2019.

\bibitem{eip2028}
Akhunov Alexey, Sasson Eli~Ben, Brand Tom, Guthmann Louis, and Levy Avihu.
\newblock {EIP}-2028: Transaction data gas cost reduction (settled).
\newblock {\em Retrieved by May 2025,
  \url{https://eips.ethereum.org/EIPS/eip-2028}}, 2019.

\bibitem{blondel2008fast}
Vincent~D Blondel, Jean-Loup Guillaume, Renaud Lambiotte, and Etienne Lefebvre.
\newblock Fast unfolding of communities in large networks.
\newblock {\em Journal of Statistical Mechanics: Theory and Experiment},
  2008(10):P10008, 2008.

\bibitem{zhou2023sok}
Liyi Zhou, Xihan Xiong, Jens Ernstberger, Stefanos Chaliasos, Zhipeng Wang,
  Ye~Wang, Kaihua Qin, Roger Wattenhofer, Dawn Song, and Arthur Gervais.
\newblock {Sok}: Decentralized finance ({DeFi}) attacks.
\newblock In {\em IEEE Symposium on Security and Privacy (SP)}, pages
  2444--2461. IEEE, 2023.

\bibitem{wang2023zero}
Zhipeng Wang, Stefanos Chaliasos, Kaihua Qin, Liyi Zhou, Lifeng Gao, Pascal
  Berrang, Benjamin Livshits, and Arthur Gervais.
\newblock On how zero-knowledge proof blockchain mixers improve, and worsen
  user privacy.
\newblock In {\em Proceedings of the ACM Web Conference (WWW)}, pages
  2022--2032, 2023.

\bibitem{gorwa2020algorithmic}
Robert Gorwa, Reuben Binns, and Christian Katzenbach.
\newblock Algorithmic content moderation: Technical and political challenges in
  the automation of platform governance.
\newblock {\em Big Data \& Society}, 7(1):2053951719897945, 2020.

\bibitem{zuo2024understanding}
Wenrui Zuo, Raul~J Mondragon, Aravindh Raman, and Gareth Tyson.
\newblock Understanding and improving content moderation in web3 platforms.
\newblock In {\em Proceedings of the International AAAI Conference on Web and
  Social Media}, volume~18, pages 1859--1870, 2024.

\bibitem{zhang2024covert}
Tao Zhang, Qianhong Wu, et~al.
\newblock Covert communication via blockchain: Hiding patterns and
  communication patterns.
\newblock {\em Computer Standards \& Interfaces (CSI)}, 90:103851, 2024.

\bibitem{zhang2023covert}
Tao Zhang, Bingyu Li, Yan Zhu, Tianxu Han, and Qianhong Wu.
\newblock Covert channels in blockchain and blockchain based covert
  communication: Overview, state-of-the-art, and future directions.
\newblock {\em Computer Communications}, 205:136--146, 2023.

\bibitem{liu2020whispers}
Shaoyuan Liu, Zhi Fang, Feng Gao, Bakh Koussainov, Zijian Zhang, Jiamou Liu,
  and Liehuang Zhu.
\newblock Whispers on {E}thereum: Blockchain-based covert data embedding
  schemes.
\newblock In {\em ACM International Symposium on Blockchain and Secure Critical
  Infrastructure (BSCI@AsiaCCS)}, pages 171--179, 2020.

\bibitem{chen2022blockchain}
Zhuo Chen, Liehuang Zhu, Peng Jiang, Can Zhang, Feng Gao, Jialing He, Dawei Xu,
  and Yan Zhang.
\newblock Blockchain meets covert communication: A survey.
\newblock {\em IEEE Communications Surveys \& Tutorials}, 24(4):2163--2192,
  2022.

\bibitem{zhu2023novel}
Liehuang Zhu, Qi~Liu, Zhuo Chen, Can Zhang, Feng Gao, and Zhongliang Yang.
\newblock A novel covert timing channel based on bitcoin messages.
\newblock {\em IEEE Transactions on Computers (TC)}, 72(10):2913--2924, 2023.

\bibitem{li2024bitcoin}
Ningran Li, Minfeng Qi, et~al.
\newblock Bitcoin inscriptions: Foundations and beyond.
\newblock {\em IEEE International Conference on Blockchain and Cryptocurrency
  (ICBC)}, 2024.

\bibitem{messias2024writing}
Johnnatan Messias, Krzysztof Gogol, Maria~In{\^e}s Silva, and Benjamin
  Livshits.
\newblock The writing is on the wall: Analyzing the boom of inscriptions and
  its impact on {EVM}-compatible blockchains.
\newblock {\em Companion of the ACM Web Conference (Workshop@WWW)}, 2025.

\bibitem{chen2023chatgpt}
Chong Chen, Jianzhong Su, Jiachi Chen, Yanlin Wang, Tingting Bi, Jianxing Yu,
  Yanli Wang, Xingwei Lin, Ting Chen, and Zibin Zheng.
\newblock When {ChatGPT} meets smart contract vulnerability detection: How far
  are we?
\newblock {\em ACM Transactions on Software Engineering and Methodology
  (TOSEM)}, 2023.

\bibitem{wang2024smartinv}
Sally~Junsong Wang, Kexin Pei, and Junfeng Yang.
\newblock Smartinv: Multimodal learning for smart contract invariant inference.
\newblock In {\em IEEE Symposium on Security and Privacy (SP)}, pages
  2217--2235, 2024.

\bibitem{sun2025adversarial}
Jiaze Sun, Zhiqiang Yin, Hengshan Zhang, Xiang Chen, and Wei Zheng.
\newblock Adversarial generation method for smart contract fuzz testing seeds
  guided by chain-based llm.
\newblock {\em Automated Software Engineering (ASE)}, 32(1):1--28, 2025.

\bibitem{ince2024detect}
Peter Ince, Xiapu Luo, Jiangshan Yu, Joseph~K Liu, and Xiaoning Du.
\newblock Detect llama-finding vulnerabilities in smart contracts using large
  language models.
\newblock {\em Australasian Conference on Information Security and Privacy
  (ACISP)}, 2024.

\bibitem{sun2024gptscan}
Yuqiang Sun, Daoyuan Wu, Yue Xue, Han Liu, Haijun Wang, Zhengzi Xu, Xiaofei
  Xie, and Yang Liu.
\newblock Gptscan: Detecting logic vulnerabilities in smart contracts by
  combining {GPT} with program analysis.
\newblock In {\em Proceedings of the IEEE/ACM International Conference on
  Software Engineering (ICSE)}, pages 1--13, 2024.

\bibitem{ma2024combining}
Wei Ma, Daoyuan Wu, Yuqiang Sun, Tianwen Wang, Shangqing Liu, Jian Zhang, Yue
  Xue, and Yang Liu.
\newblock Combining fine-tuning and {LLM}-based agents for intuitive smart
  contract auditing with justifications.
\newblock {\em Proceedings of the IEEE/ACM International Conference on Software
  Engineering (ICSE)}, 2025.

\bibitem{ge2025adapting}
Huilin Ge, Ze~Wang, Runbang Liu, Zhiwen Qiu, Jie Xia, Ting Chen, and Hongzi
  Zhu.
\newblock Adapting large language models for smart contract defects detection
  in the open network blockchain.
\newblock {\em IEEE Internet of Things Journal (IOTJ)}, 2025.

\bibitem{xiao2025logic}
ZeKe Xiao, Qin Wang, Hammond Pearce, and Shiping Chen.
\newblock Logic meets magic: {LLM}s cracking smart contract vulnerabilities.
\newblock {\em IEEE International Conference on Blockchain and Cryptocurrency
  (ICBC)}, 2025.

\bibitem{li2025scalm}
Zongwei Li, Xiaoqi Li, Wenkai Li, and Xin Wang.
\newblock {SCALM}: Detecting bad practices in smart contracts through {LLM}s.
\newblock {\em The Association for the Advancement of Artificial Intelligence
  (AAAI)}, 2025.

\bibitem{lei2025large}
Yuchen Lei, Yuexin Xiang, et~al.
\newblock Large language models for cryptocurrency transaction analysis: A
  {B}itcoin case study.
\newblock {\em arXiv preprint arXiv:2501.18158}, 2025.

\bibitem{avice2025linking}
R{\'e}gnier Avice, Bernhard Haslhofer, Zhidong Li, and Jianlong Zhou.
\newblock Linking cryptoasset attribution tags to knowledge graph entities: An
  {LLM}-based approach.
\newblock {\em Financial Cryptography and Data Security (FC)}, 2025.

\bibitem{luo2025decoding}
Junliang Luo, Xihan Xiong, William Knottenbelt, and Xue Liu.
\newblock Decoding {SEC} actions: Enforcement trends through analyzing
  blockchain litigation using {LLM}-based thematic factor mapping.
\newblock {\em The International Conference on Artificial Intelligence and Law
  (ICAIL)}, 2025.

\bibitem{liu2024propertygpt}
Ye~Liu, Yue Xue, Daoyuan Wu, Yuqiang Sun, Yi~Li, Miaolei Shi, and Yang Liu.
\newblock Property{GPT}: {LLM}-driven formal verification of smart contracts
  through retrieval-augmented property generation.
\newblock {\em Network and Distributed System Security (NDSS) Symposium}, 2025.

\bibitem{jiang2024unearthing}
Jinan Jiang, Zihao Li, Haoran Qin, Muhui Jiang, Xiapu Luo, Xiaoming Wu, Haoyu
  Wang, Yutian Tang, Chenxiong Qian, and Ting Chen.
\newblock Unearthing gas-wasting code smells in smart contracts with large
  language models.
\newblock {\em IEEE Transactions on Software Engineering (TSE)}, 2024.

\bibitem{han2021trust}
Jianlei Han, Shiyang Huang, and Zhuo Zhong.
\newblock Trust in defi: an empirical study of the decentralized exchange.
\newblock {\em Available at SSRN}, 3896461, 2021.

\bibitem{wu2025hunting}
Zhiying Wu, Jiajing Wu, Hui Zhang, Zibin Zheng, and Weiqiang Wang.
\newblock Hunting in the dark forest: A pre-trained model for on-chain attack
  transaction detection in web3.
\newblock In {\em Proceedings of the ACM on Web Conference (WWW)}, 2025.

\bibitem{daian2020flash}
Philip Daian, Steven Goldfeder, Tyler Kell, Yunqi Li, Xueyuan Zhao, Iddo
  Bentov, Lorenz Breidenbach, and Ari Juels.
\newblock Flash boys 2.0: Frontrunning in decentralized exchanges, miner
  extractable value, and consensus instability.
\newblock In {\em IEEE symposium on security and privacy (SP)}, pages 910--927,
  2020.

\bibitem{wang2023brc}
Qin Wang and Guangsheng Yu.
\newblock {BRC-20}: Hope or hype.
\newblock {\em arXiv preprint arXiv:2310.10652}, 2023.

\bibitem{yang2024stole}
Jingjing Yang, Jieli Liu, Dan Lin, Jiajing Wu, Baoying Huang, Quanzhong Li, and
  Zibin Zheng.
\newblock Who stole my {NFT}? investigating web3 {NFT} phishing scams on
  {E}thereum.
\newblock {\em IEEE Transactions on Information Forensics and Security (TIFS)},
  2024.

\bibitem{bolz2024machine}
Manuel Bolz, Kevin Br{\"u}ndler, Liam Kane, Panagiotis Patsias, Liam
  Tessendorf, Krzysztof Gogol, Taehoon Kim, and Claudio Tessone.
\newblock Machine learning-based detection of pump-and-dump schemes in
  real-time.
\newblock {\em IEEE International Conference on Blockchain and Cryptocurrency
  (ICBC)}, 2025.

\bibitem{wang2024cryptocurrency}
Qin Wang, Guangsheng Yu, and Shiping Chen.
\newblock Cryptocurrency in the aftermath: Unveiling the impact of the {SVB}
  collapse.
\newblock {\em IEEE Transactions on Computational Social Systems (TCSS)}, 2024.

\bibitem{eip4444}
Kadianakis George, lightclient, and Stokes Alex.
\newblock {EIP}-4444: Bound historical data in execution clients (stagnant).
\newblock {\em Retrieved by May 2025,
  \url{https://eips.ethereum.org/EIPS/eip-4444}}, 2021.

\bibitem{zhou2024llm}
Ruiyang Zhou, Lu~Chen, and Kai Yu.
\newblock Is {LLM} a reliable reviewer? a comprehensive evaluation of {LLM} on
  automatic paper reviewing tasks.
\newblock In {\em Proceedings of the Joint International Conference on
  Computational Linguistics, Language Resources and Evaluation}, pages
  9340--9351, 2024.

\bibitem{zhong2024can}
Li~Zhong and Zilong Wang.
\newblock Can {LLM} replace stack overflow? a study on robustness and
  reliability of large language model code generation.
\newblock In {\em Proceedings of the AAAI Conference on Artificial Intelligence
  (AAAI)}, volume~38, pages 21841--21849, 2024.

\end{thebibliography}

\appendix

\section{LLM Prompt Example}
\label{appendix: prompt}

\begin{figure*}[t]
\centering
\begin{lstlisting}[style=custompython,caption={LLM prompt for IDM emotion classification.}]
def analyze_IDM_sentiment_using_LLM(message: str):
    prompt = f"""
    (*@\textbf{You are tasked with}@*) analyzing blockchain input data messages (IDMs), which may include both structured tokens (e.g., wallet addresses, transaction hashes) and natural language content.
    (*@\textbf{Your objective}@*) is to extract the dominant human emotion conveyed in the natural language portion of the message.
    
    (*@\textbf{Instructions}:@*)
    1. (*@\textcolor{blue!70}{Focus on the natural language content. Disregard structured tokens unless they contribute to the emotional tone.}@*)
    2. (*@\textcolor{blue!70}{Identify the most appropriate emotion label (e.g., 'Positive-Joy') using the taxonomy provided below.}@*)
    3. (*@\textcolor{blue!70}{For that label, also provide:}@*)
       - (*@\textcolor{blue!70}{An intensity score (1--10), reflecting the strength of emotional expression.}@*)
       - (*@\textcolor{blue!70}{A confidence score (0.0--1.0), reflecting your certainty in the label assignment.}@*)
    (*@\textbf{Note}:@*) When evaluating emojis, consider them only if they support coherent emotional meaning in context. Ignore isolated or repetitive emoji sequences lacking semantic relevance.

    ### Emotive taxonomy with definitions:
    - (*@\textcolor{violet}{Positive-Joy}@*): Expressions of happiness, celebration, or emotional uplift.
    - (*@\textcolor{violet}{Positive-Trust}@*): Expressions of belief, confidence, or endorsement toward a person, group, or project.
    - (*@\textcolor{violet}{Positive-Warmth}@*): Light, friendly, and kind expressions, such as greetings, farewells, or general goodwill.
    - (*@\textcolor{violet}{Positive-Love}@*): Strong emotional attachment, romantic or platonic affection, and heartfelt personal expression.
    - (*@\textcolor{violet}{Positive-Hope}@*): Positive expectations, future-oriented optimism, or messages expressing wishes or aspirations.
    - (*@\textcolor{violet}{Positive-Gratitude}@*): Appreciation, praise, thanks, or motivational support toward others.
    - (*@\textcolor{violet}{Negative-Anger}@*): Strong emotional reaction of frustration or outrage, often over perceived injustice.
    - (*@\textcolor{violet}{Negative-Hostility}@*): Targeted verbal aggression, threats, insults, or confrontational language.
    - (*@\textcolor{violet}{Negative-Sadness}@*):  Expressions of grief, loss, hopelessness, or emotional pain.
    - (*@\textcolor{violet}{Negative-Fear}@*): Messages that express anxiety, suspicion, or caution about risks, scams, or vulnerabilities.
     - (*@\textcolor{violet}{Negative-Disgust}@*): Moral/emotional revulsion, often directed at people/blockchain projects in our context.
    - (*@\textcolor{violet}{Negative-Guilty}@*): Apologies, self-blame, regret, or acknowledgment of fault or failure.
    - (*@\textcolor{violet}{Neutral-Surprise}@*): Statements of astonishment.
    - (*@\textcolor{violet}{Neutral-Confusion}@*): Expressions of uncertainty, lack of understanding, or requests for clarification.
    - (*@\textcolor{violet}{Neutral-Politeness}@*): Routine formalities or courteous phrases such as greetings, closings, or expressions of etiquette, low in emotional intensity.
    - (*@\textcolor{violet}{Neutral-Curiosity}@*): Genuine or exploratory questions reflecting a desire to understand or inquire.

   (*@\textcolor{blue!70}{If no human emotion is detected, assign a label from the non-emotive taxonomy below.}@*)
    
    ### Non-emotive taxonomy with definitions:
    - (*@\textcolor{violet}{Functional-Technical}@*): Functional messages like code, transaction record, or deployment notices without emotional intent.
    - (*@\textcolor{violet}{Functional-Operational}@*): On-chain activity logs or declarative statements describing state changes or system behavior, without subjective emotion or human-directed intent.
    - (*@\textcolor{violet}{Uninterpretable-Unclear}@*): Partially readable or human-like messages that are semantically ambiguous, fragmented, or incoherent, making emotional interpretation unreliable.
    - (*@\textcolor{violet}{Uninterpretable-Garbage}@*): Non-linguistic or noise-like content such as long base64 strings, hex data, emoji floods, or meaningless character sequences, with no interpretable intent or emotion.

    (*@\textbf{Output format}:@*) (*@\textcolor{blue!70}{Return a list in the format: \texttt{[label, intensity, confidence] }}@*)
    - For (*@\textcolor{violet}{emotive}@*) messages: provide a label from the emotive taxonomy, an intensity score (1--10), and a confidence score (0.0--1.0).
    - For (*@\textcolor{violet}{non-emotive}@*) messages: provide a label from the non-emotive taxonomy, set intensity to `null`, and include the confidence score.

    (*@\textcolor{blue!70}{Now analyze the following message}:@*)
    \"\"\"{message}\"\"\"
    """
    
    response = client.chat.completions.create(
        model = "openai/gpt-4o-2024-11-20",
        messages = [ {"role": "user", "content": prompt} ], temperature = 0.2)
    
    return response.choices[0].message.content
\end{lstlisting}
\end{figure*}

\end{document}